\documentclass[useAMS,usenatbib,12pt,preprintnumbers,nofootinbib,onecolumn,notitlepage]{revtex4-2}
\usepackage[T1]{fontenc} 
\usepackage{graphicx}
\usepackage{subcaption}
\usepackage{amsmath} 
\usepackage{verbatim} 
\usepackage{mathrsfs}
\usepackage{amsfonts}
\usepackage{color}
\usepackage{hyperref}
\usepackage{caption}
\newcommand{\nk}{\textbf{k}}

\newcommand{\x}{\textbf{x}}
\newcommand{\y}{\textbf{y}}
\newcommand{\bra}{\langle}
\newcommand{\ket}{\rangle}
\newcommand{\mH}{\mathcal{H}}
\newcommand{\mP}{\mathcal{P}}
\newcommand{\mR}{\mathcal{R}}

\newcommand{\mT}{\mathcal{T}}
\newcommand{\mM}{\mathcal{M}}
\newcommand{\nn}{\nonumber \\}

  \newcommand{\ini}{\text{ini}}
  \newcommand{\eq}{\text{eq}}
 \newcommand{\fin}{\text{end}}

\begin{document}

\title{Inflation and the cosmological (not-so) constant in unimodular gravity}

\author{Gabriel Le\'{o}n}
\email{gleon@fcaglp.unlp.edu.ar }
\affiliation{Grupo de Cosmolog\'{\i}a, Facultad
	de Ciencias Astron\'{o}micas y Geof\'{\i}sicas, Universidad Nacional de La
	Plata, Paseo del Bosque S/N 1900 La Plata, Argentina.\\
	CONICET, Godoy Cruz 2290, 1425 Ciudad Aut\'onoma de Buenos Aires, Argentina. }

\begin{abstract}

We propose a mechanism for generating an inflationary phase in the early universe without resorting to any type of scalar field(s).  Instead, this accelerated expansion is driven by a dynamical ``cosmological constant'' in the framework of unimodular gravity.   The time dependent cosmological constant can be related  to an energy diffusion term that arises naturally in unimodular gravity due to its restrictive diffeomorphism  invariance.  We derive the  generic conditions required for any type of diffusion to generate a realistic  inflationary epoch. Furthermore, for a given parameterization of inflation (in terms of the Hubble flow functions), we show how to construct the corresponding diffusion term in such a way that a smooth transition occurs between inflation and the subsequent radiation dominated era, hence reheating proceeds naturally.  The primordial spectrum is obtained during the inflationary phase by considering inhomogeneous perturbations associated to standard hydrodynamical matter (modeled as a single ultra-relativistic fluid). We demonstrate that the resulting spectrum is equivalent to that obtained in traditional inflationary models, and is also independent of the particular form of the diffusion term.  In addition,  we analyze the feasibility of identifying  the variable cosmological constant, responsible for the inflationary expansion,  with the current observed value.

\end{abstract}


\maketitle

\section{Introduction}

The standard $\Lambda$CDM cosmological model, along with the inflationary paradigm \cite{Guth81,Hawking82,Linde82,Linde83,Mukhanov81,mukhanov92},  has proven to be very  successful in describing the most recent cosmological observations such as the Cosmic Microwave Background (CMB) \cite{Planck18a,Planck18b,Planck18c},  Type Ia supernovae \cite{Scolnic18}, and baryon acoustic oscillations \cite{eBOSS}.  The model is characterized by a spatially flat, expanding universe whose dynamics are governed by Einstein Field Equations (EFE) of General Relativity. The matter constituents are dominated by cold dark matter (CDM) and a cosmological constant (CC)  $\Lambda$ at late times. The inclusion of $\Lambda$ is equivalent to the introduction of a uniformly distributed form of energy, usually referred to as ``dark energy'', which is responsible for the late time accelerated expansion in the universe \cite{Riess98,Perlmutter99,Abbott1811}.  Additionally, during the early stages in the evolution of the universe, an accelerated expansion is also assumed to have taken place.\footnote{Throughout this work, we employ the term  ``early accelerated expansion'',  to refer to the inflationary phase. Namely, do not confuse it with \textit{early dark energy} models.} The primordial seeds of cosmic structure are generally attributed to  vacuum fluctuations of the scalar field, the inflaton, which drives the early accelerated expansion known as inflation.

In spite of the success of the $\Lambda$CDM model, there are some elements that have not been completely understood. For instance, there is little knowledge about the precise nature of the dark sector, i.e. dark matter and dark energy. Concerning the latter,  from the theoretical point of view,  examinations within quantum field theory  lead to the so-called CC problem \cite{Weinberg89,Carroll92,JMartin12,Rugh00,Bengochea2019}. The issue consists in a vast  disagreement (between 50 and 120 orders of magnitude) between the small observed value of the CC and the large theoretical prediction based on the quantum vacuum energy density,  which is supposed to act as a CC. As a result,
in order to account for the observed value, one must rely on arguments involving extreme fine-tuning.  This, in short, is the (old) CC problem that has motivated some specialists to abandon the constant $\Lambda$ and use a dynamical field to explain the current value of the CC. The most popular dynamical fields include quintessence \cite{quintessence1,quintessence2,quintessence3,chinoEDE1,chinoEDE2,chinoEDE3}, phantom  \cite{phantom} and quintom \cite{quintom1,quintom2,quintom3,quintom4}.  In addition, modified gravity theories have been widely used to replace $\Lambda$ see e.g.  \cite{Clifton2011,Bengochea2008,Bengochea2010,celia2021,Nojiri2010}.

In regards to the early accelerated expansion of the universe, there are some aspects that are not entirely comprehended either. As is well known, General Relativity (GR) in presence of standard fluids cannot describe a stage of accelerated expansion. In the case of the early universe, this led to the introduction of the inflaton: a scalar field, which under some  circumstances, can trigger the  cosmological inflation. However,   the naturalness of the inflationary paradigm has  been criticized before \cite{Ijjas2013,penrose2016fashion,Gibbons2006},  questioning, in particular, the special initial conditions required for inflation to actually begin (see however \cite{jmartin2019}). 
Some additional issues we can mention are: the subject of eternal inflation, a feature that is present in almost every model of inflation \cite{Vilenkin1983}, which leads to the controversial topic of the multiverse;  and the trans-Planckian problem for primordial perturbations \cite{jmartin2000}.  As a consequence,  different models of inflation, in which the expansion is not driven by the inflaton, have been considered by  specialists, see e.g.  \cite{Wald2002,Watson2006,Bengochea2014,luisa2018,luisa2021,Jaime2022}.

In view of the above arguments, it is reasonable to explore alternatives to the $\Lambda$CDM model that may account for some of the problems mentioned. Unimodular Gravity (UG) is an alternative approach to GR that  can be derived from the Einstein-Hilbert action by restricting to variations preserving the volume element \cite{UG0,UG1,UG2,UG3,UG4,ellis2010}; hence, UG has less symmetries than GR.  While the  implementation of UG into the cosmological context, and its connection with the CC, has been noted for decades \cite{Weinberg89,ellis2010,UG4,Smolin2009,Uzan2010}, recently it has been rediscovered, gaining significant attention \cite{ellis2015,Corral20,Nucamendi20,sudarskyPRL1,sudarskyPRL2,sudarskyH0,Daouda2018,miguelangel1,miguelangel2,GUG1,GUG2,deCesare2021,Nojiri2015,Nojiri2016}. 

The gravitational field in UG is characterized by the trace-free Einstein equations, which is a subset of the EFE that can give back the full EFE with an integration
constant. This is due to the fact that in UG conservation of the energy-momentum tensor is not a necessary assumption of the theory. Namely, it is considered as an extra hypothesis that, when imposed, the cosmological constant $\Lambda_*$ in EFE arises as an integration constant and then is fixed by initial data \cite{ellis2010,ellis2015}.  In this way, the vacuum energy density is now unrelated to $\Lambda_*$, so is removed from the gravitational field equations. In other words, vacuum energy does not gravitate in UG \cite{Weinberg89}.  Therefore, one eliminates the (old) CC problem; nonetheless,  UG--by itself--does not offer an explanation for the magnitude of the CC inferred from astronomical observations.

An interesting take on the subject, originally proposed in \cite{sudarskyPRL1} is to adopt the non-conservation of the energy-momentum tensor and use it to characterize a dynamical CC.  The time-dependent CC  records the cumulative effect of such non-conservation resulting in an effective \textit{energy diffusion term}.  In \cite{sudarskyPRL1,sudarskyPRL2} it was argued that the micro-physical origin of the diffusion term could be traced back to quantum gravitational effects. Specifically,  fundamental spacetime discreteness could lead to small violations of translational invariance (for example 
in the context of causal sets \cite{Dowker2004,Dowker2009}).  Henceforth, according to \cite{sudarskyPRL1,sudarskyPRL2}, the  origin of the CC might be found in the microscopic structure of spacetime and its interaction with matter. Under these considerations, the authors of \cite{sudarskyPRL1,sudarskyPRL2} provided an estimate of the order of magnitude of the present CC with positive results.  It is worth mentioning some differences, regarding the physical nature of the dynamical CC, between the previous approach (due to UG) and the better known one based on a gravity-coupled scalar field (e.g. quintessence or slow-roll inflation).  In the former approach (as we have mentioned) the energy diffusion term could be tied to a fundamental description of  spacetime. On the other hand, when considering a scalar field(s) coupled to gravity, one needs to specify the nature of such a field(s).  While fundamental particle physics theories, such as string theory implementations, can account for the existence of several kinds of scalar fields, the fact is that currently the only fundamental scalar field observed in Nature is the Higgs boson.  Therefore, in order to produce an accelerated cosmic expansion within the latter framework  (QFT plus GR), one requires to postulate the existence of one (or more) fundamental scalar field(s) besides the Higgs field.  Moreover, one needs to specify the characteristics of the corresponding potential term, which normally implies the introduction of several free parameters.\footnote{This is generally true if one wants to unify the inflationary epoch with the later accelerated expansion, see e.g. \cite{chinoEDE1,chinoEDE2,chinoEDE3}.  }   In contrast,  a dynamical CC characterized by an effective energy diffusion term, whose underlying origin might come from quantum gravitational effects, seems an alternative proposal that deserves to be explored.  In particular, if the same diffusion term can address simultaneously the early inflationary epoch and the later accelerated expansion  using a minimal set of assumptions (and parameters).  

Following the aforementioned approach, in Ref.  \cite{aperez2021} the energy-momentum non-conservation due to a fundamental granularity of the spacetime at Planckian scales, was used to characterize an inflationary phase. Furthermore, in that same work, the primordial spectrum was generated by resorting only to fluctuations of the Higgs scalar field during the inflationary regime, within the semiclassical gravity framework. The resulting spectrum was nearly scale invariant with the right amplitude.  

In the present work, we will also accept the premise of a dynamical CC, within the UG framework, effectively characterized by a diffusion term.\footnote{We invite the reader to consult Refs. \cite{Elias2020,Hermano2021}, for a review of proposals discussing the theoretical and conceptual constructions surrounding the non-conservation of the energy-momentum tensor.} In general terms, our aim is to use such a hypothesis to model a realistic inflationary regime and investigate its possible connection with the observed value of the CC.  The main characteristics of a reasonable inflationary epoch are:  it must last long enough (in order to solve the horizon problem \cite{BaumannTASI2009}) and exhibit a graceful exit into an era dominated by radiation.  The main differences between our work and Ref. \cite{aperez2021} is that we will not employ a particular scalar field for generating the primordial inhomogeneities nor will we resort to semiclassical gravity. Additionally, we will not be concerned with the micro-physical origin of the diffusion term. Instead, we will derive the conditions under which a generic diffusion function can produce an inflationary phase.\footnote{Although we will not deal with the micro-physics at stake, it is worth mentioning that the dynamical evolution, obtained in the present work, corresponding to the diffusion term and the Hubble radius during inflation is very similar to that of Ref. \cite{aperez2021}. See Secs. \ref{sec_inf} and \ref{sec_CC}. }  This approach has the same spirit as the usual method of modeling slow-roll inflation.  In the latter, one finds the conditions required for the inflaton potential (e.g. the slow-roll conditions) to produce an accelerated expansion, but without specifying the high energy theory that leads to a specific potential.

 In regards to the matter content in the early epoch, we will assume that, since the beginning of inflation, matter can  always be modeled by a single perfect fluid with an equation of state corresponding to pure radiation, that is to say,   matter behaves as an ultra-relativistic fluid during the early expansion.  Moreover, we will employ cosmological perturbation theory in UG and apply it directly to the density (of the radiation fluid) and metric perturbations. As we will show in Sec. \ref{sec_PS}, the familiar Mukhanov-Sasaki equation is retrieved for any generic diffusion term as long as it is homogeneous. Consequently, the predicted power spectrum  is completely equivalent to that obtained from traditional inflationary models (e.g. slow-roll inflation).   In order to perform a quantitative analysis, we will construct a particular form of the diffusion function compatible with a realistic inflationary regime, parameterized through the Hubble flow functions \cite{terreroHFF,terreroHFF2}. For such constructed diffusion term, we analyze its dynamical evolution until the present epoch, and show that the estimated order of magnitude is similar to the observed value of the CC. 

The article is organized as follows:  In Sec. \ref{seccosmoUG}, we provide a very brief review of unimodular gravity, and focus on its implementation into the cosmological context. In Sec. \ref{sec_inf},  we find the conditions required for a generic diffusion term to generate an inflationary phase. Additionally, we show how a specific diffusion term can be constructed from a general parameterization of inflation in UG. In Sec. \ref{sec_PS}, we make use of cosmological perturbation theory in UG to calculate the primordial power spectrum. In Sec. \ref{sec_CC}, we analyze how the observed value of the cosmological constant can be identified with the variable CC that is also responsible for inflation. We also provide estimates for other cosmological observables of interest.  Lastly in Sec. \ref{conclusions}, we present our conclusions. Regarding conventions, we use the metric signature $(-,+,+,+)$ and units where $c=\hbar=1$;  the Riemann tensor is $R^\alpha_{\: \beta \mu \nu} = \partial_\mu \Gamma^\alpha_{\beta \nu} + \ldots $; the Ricci tensor is $R_{\mu \nu} = R^\lambda_{\: \mu \lambda \nu}$ and $R = g^{\mu \nu} R_{\mu \nu}$.

\section{Cosmological equations  in unimodular gravity} \label{seccosmoUG}

There are several equivalent formulations of unimodular gravity  \cite{UG0,UG1,UG2,UG3,UG4}. A possible realization is given by writing the Einstein-Hilbert action and introduce the unimodular condition as a Lagrange multiplier
\begin{eqnarray}\label{accion0}
	S[g_{\mu \nu}, \lambda, \varphi] &=& \frac{M_P^2}{2} \int d^4x \sqrt{-g} \left[   R    - 2 \lambda(x) \left( 1 - \frac{\varepsilon_0(x)}{\sqrt{-g}}  \right)   \right]   \nn
	&+&  S_m [g_{\mu \nu},\varphi],
\end{eqnarray}
where $M_P^2 \equiv 1/(8 \pi G)$ is the reduced Planck mass, $S_m$ is the action corresponding to the matter fields, which are represented by $\varphi$, and $\varepsilon_0 (x)$ is a non-dynamical 4-from that breaks down the diffeomorphism symmetry of GR to  volume preserving diffeomorphisms, which restricts the symmetries of traditional GR.  

Varying  action \eqref{accion0} with respect to the (inverse) metric, leads to
\begin{equation}\label{EE1}
	R_{\mu \nu} - \frac{1}{2} g_{\mu \nu} R + \lambda(x) g_{\mu \nu} = \mathcal{T}_{\mu \nu},
\end{equation}
where $\mathcal{T}_{\mu \nu} \equiv  T_{\mu \nu}/M_P^2$ and $T_{\mu \nu} = -2 (-g)^{-1/2} \delta S_m/ \delta g^{\mu \nu}$ is the energy-momentum tensor of the matter fields.  Equations \eqref{EE1} can be interpreted as Einstein's field equations  (EFE) with an effective ``cosmological constant''  $\lambda(x)$; note that the volume element of the metric $g_{\mu \nu}$ is fixed through the condition $\varepsilon_0(x) = \sqrt{-g}$.

Taking the trace in Eq. \eqref{EE1} yields
\begin{equation}\label{lambda}
	\lambda(x) = \frac{1}{4} (R + \mathcal{T} ).
\end{equation}
Replacing $\lambda$ back into  Eq. \eqref{EE1}, one obtains
\begin{equation}\label{EEtracefree}
R_{\mu \nu} - \frac{1}{4} g_{\mu \nu} R = 	\mathcal{T}_{\mu \nu} - \frac{1}{4} g_{\mu \nu} \mathcal{T}.
\end{equation}
These are the trace-free Einstein field equations and are a new set of equations for the gravitational field. The theory characterized by Eqs. \eqref{EEtracefree} has been dubbed Unimodular Gravity (UG). 

The volume preserving diffeomorphisms, restricting the symmetry of the Einstein-Hilbert action, can be generated by a vector field $\chi^\nu$ satisfying $\nabla_\nu \chi^\nu = 0$. The latter equation is solved as $\chi^\nu = (1/2) \epsilon^{\nu \alpha \beta \gamma} \nabla_\alpha \omega_{\beta \gamma}$ where $\omega_{\beta \gamma}$ is an arbitrary two-form. This reduced symmetry in the action implies, by Noether's theorem,  that \cite{sudarskyPRL1,aperez2021}
\begin{equation}\label{conservEM}
	\nabla^\mu (T_{\mu \nu} - g_{\mu \nu} Q) = 0,
\end{equation}
where $Q(x)$ is  an arbitrary function that quantifies the violation of the conservation of  $T_{\mu \nu}$, from now on, we will refer to $Q$ as the \textit{diffusion term}. On the other hand, if one fixes $Q=$ constant, the usual conservation law is recovered. Therefore, conservation of $T_{\mu \nu}$ is introduced in UG as an additional premise.  In this work, we will drop this assumption as a varying $Q$ will play an important role in our model. 

Applying $\nabla^\mu$ on both sides of Eq. \eqref{EE1}, together with Eq. \eqref{conservEM}, leads to $\nabla_\mu \lambda(x) =  M_P^{-2} \nabla_\mu  Q(x)$, which can be solved as
\begin{equation}\label{lambdaeff}
	\lambda(x) = \Lambda_* + \frac{Q}{M_P^2}.
\end{equation}
In this case $\Lambda_*$ is an integration constant fixed by the initial data. In fact, as is well known \cite{Weinberg89,ellis2010,ellis2015}, assuming the conservation of $T_{\mu \nu}$,  implies that $\lambda(x)$ is a true (cosmological) constant whose value is fixed by observations. However, if $Q(x)$ is not a constant, then one has a variable ``cosmological constant'' sourced by non-conservation of the energy-momentum tensor.  

Our next task is to apply the UG theory into the cosmological context.  Assuming the cosmological principle, and spatial flatness, the line element is given by the Friedmann-Lemaitre-Robertson-Walker (FLRW) metric
\begin{equation}\label{metricaFRW}
	ds^2 = -dt^2  + a^2 (t) (dx^2 + dy^2 + dz^2).
\end{equation}
The UG constraint, resulting from the previous form of the FLRW metric, is given as $\varepsilon_0 = a^3$.
In addition, the assumption that the universe is spatially flat can be justified since this is consistent with the most recent observational data (see e.g. \cite{Planck18a}). Furthermore, in the next section we will show  how  an inflationary phase can be obtained. Consequently, one expects that inflation will erase any spatial curvature that might be initially present, provided that inflation lasts long enough. 

We  characterize the matter content in the universe by the energy-momentum tensor of a perfect fluid,
\begin{equation}\label{EMtensor}
	T_{\mu \nu} = (\rho + p ) u_\mu u_\nu + p g_{\mu \nu},
\end{equation}
where $\rho$ and $p$ are the energy density and pressure in the rest frame of the fluid respectively, and $u^\mu$  represents its 4-velocity (relative to the observer) normalized as $u^\mu u_\mu = -1 $.  Moreover,  we assume that the energy-momentum tensor and the source $Q$ are also spatially  isotropic and homogeneous, i.e. $\rho(t)$, $p(t)$ and $Q(t)$.  These assumptions yield the set of Friedmann's equations
\begin{equation}\label{F1}
3	H^2 = \frac{1}{ M_P^2} (\rho + Q) + \Lambda_*,
\end{equation}
\begin{equation}\label{F2}
	2 \dot H + 3 H^2 = \frac{1}{M_P^2} (-p + Q) + \Lambda_*,
\end{equation}
where $\rho,p$ represent the total energy density and pressure of the matter content in the universe respectively;  the dot denotes derivative with respect to cosmic time $t$ and $H \equiv \dot a/a$ is the Hubble factor.  The  continuity equation of the energy-momentum tensor, Eq. \eqref{conservEM}, reads
\begin{equation}\label{rhomov}
\dot \rho + \dot Q + 3 H(\rho + p) = 0.
\end{equation}

As we have mentioned, the integration constant $\Lambda_*$ can be fixed by initial data. Using Eqs. \eqref{lambda}, \eqref{lambdaeff}, together with the  FLRW metric \eqref{metricaFRW} and  $T_{\mu \nu }$ in  Eq. \eqref{EMtensor}, one finds
\begin{equation}\label{lambdaini}
\Lambda_*	= - \frac{1}{ M_P^2} (\rho_\ini + Q_\ini)  + 3	H_\ini^2.
\end{equation}
The previous equation can also be found directly from Eq. \eqref{F1}.

\section{The background: conditions for  the inflationary phase}\label{sec_inf}

In this section we will use the cosmological equations derived in Section \ref{seccosmoUG}, and explore the relation between  $Q$ and inflation.  Criteria based on naturalness of initial conditions suggest the initial values 
\begin{equation}\label{condini}
	\rho^\ini \simeq Q^\ini \simeq M_P^4,
\end{equation}
right at the beginning of the expansion. Therefore, we can fix the value of the integration constant as $\Lambda_* = 0$. This choice implies, from Eq. \eqref{lambdaini}, that $3 H^{\ini 2} \simeq 2 M_P^2$. 

In order to parameterize the full background dynamics, including the inflationary regime, it is convenient to introduce the Hubble flow functions (HFF) \cite{terreroHFF,terreroHFF2} defined as
\begin{equation}\label{HFF}
	\epsilon_{n+1} =\frac{ \dot \epsilon_{n}}{H \epsilon_{n}}, \qquad \epsilon_{0}=\frac{H^\ini}{H},
\end{equation} 
where $n =0, 1, 2, \ldots$  In particular, the first HFF is expressed as  $\epsilon_1 = -\dot H/{H^2}$.  From Friedmann's equations \eqref{F1} \eqref{F2}, and  assuming the equation of state (EOS) for the perfect fluid as $p = w \rho$, one obtains an expression for the first HFF,
\begin{equation}\label{eps1gen}
	\epsilon_1 =  \frac{3(1 + w)}{2(1+\Gamma)},  \qquad   \text{where}  \qquad  \Gamma \equiv Q/\rho.
\end{equation} 
We observe from  Eq. \eqref{eps1gen}, that if $\Gamma$ satisfies the condition $\Gamma > (1+3w)/2$, then $\epsilon_{1} < 1$, meaning that an inflationary phase can take place.  As we will show shortly, a similar condition exists in traditional inflation models.



In this work,  we will assume that,  from the beginning of the inflationary expansion, the total matter content of the universe  behaves as a hydrodynamical fluid consisting of pure radiation.  
Therefore, we assume an EOS $p = w \rho$ with $w=1/3$  valid from the beginning of inflation, during inflation, and up to near the end of the radiation dominated epoch (we will be more precise in Sec. \ref{sec_CC}). For $w=1/3$, the continuity equation \eqref{rhomov} is expressed as
\begin{equation}\label{rhomovN}
	\rho(N),_N + Q(N),_N +4 \rho(N) =0,
\end{equation}
where $ f,_N$ denotes derivative of $f$ with respect to  $N$. Also, we have performed a change of variable $H dt = dN$, where $N$ is the number of e-folds,  defined as $a(N) \equiv e^{N} a_\ini$.  The analytical solution to Eq.  \eqref{rhomovN},  for a given $Q(N)$ satisfying initial conditions \eqref{condini}, is  
\begin{equation}\label{rhosolrad0}
	\rho (N)= M_P^4 e^{-4 N} - e^{-4 N} \int_{0}^N e^{4 \bar{N}} Q,_{\bar N} d\bar N.
\end{equation}

In addition,  for $w=1/3$,  the first HFF \eqref{eps1gen} takes the form
\begin{equation}\label{epsilon1}
	\epsilon_1 = \frac{2}{ 1 + \Gamma}.
\end{equation}

Let us describe the expected behavior of $\epsilon_1$ during inflation. 
The (natural) initial conditions chosen previously \eqref{condini},  imply the initial value $\epsilon_1^\ini \equiv \epsilon_1(0)= 1$ (since $\Gamma(0) \simeq 1$). Inflation can occur generically  if $\epsilon_1 < 1$.  In particular, if $|\epsilon_n| \ll 1$ for any $n$, one would have an equivalent accelerated expansion as characterized by standard slow-roll inflationary models.  Additionally, after a sufficient time, inflation must end and  a graceful exit towards a radiation dominated era must ensue.  This transition can be characterized by  identifying $\epsilon_1$ with a monotonic increasing function; when its value reaches up to $ \epsilon_1 \simeq 1$, inflation ends.  And a further increase to $\epsilon_1 \lesssim  2$  ensures the beginning of the radiation dominated period.  

We are now in a position to find the specific conditions that the diffusion term $ Q $ must satisfy to be compatible with an inflationary phase.  As we have mentioned, inflation occurs if $\epsilon_1 < 1$, this is equivalent, by Eq. \eqref{epsilon1},  to the condition
\begin{equation}\label{condicionQinf}
	\frac{Q}{\rho} > 1.
\end{equation}
Namely, if the diffusion term $Q$ dominates over $\rho$ (with EOS parameter $w=1/3$), then inflation is assured.  This is tantamount to the condition used in traditional slow-roll inflationary models,  where  one imposes that the inflaton  potential must dominate over the kinetic term for inflation to take place.\footnote{ In terms of the pressure and energy density, the first HFF reads $\epsilon_1 = 3(\rho+p)/2\rho$.  Assuming that inflation is driven by a single homogeneous scalar field $\varphi(t)$,  then $\epsilon_1 =3/(1+2 V/\dot \varphi^2)$. So inflation is achieved by requiring that the potential  dominates over the kinetic term, i.e. $V \gg \dot \varphi^2/2$.  } Here it is important to mention that, since the ultra-relativistic fluid is always present, and both $Q$ and $\rho$ start close to Planckian scales,  quantum effects could induce fluctuations in the energy density large enough such that  a back-reaction occurs, affecting the expansion. For instance, it could happen that quantum fluctuations produce an effective $\rho_\text{eff.}$  that oscillates around $Q$. If that were the case, then one would have various micro-phases of accelerated and decelerated expansion in a very short interval of e-folds. Therefore, condition  \eqref{condicionQinf} is not trivial to assure without considering the quantum effects of the (ultra-relativistic) matter fields,  and  possibly of the diffusion term.   However, we will proceed with the analysis assuming that  Eq. \eqref{condicionQinf}  is valid long enough, at least for a minimum of $N_\text{min.}\simeq 10^2$ e-folds, to generate an inflationary phase. The condition to end inflation is given by $Q \simeq \rho$.    And the condition $Q \lesssim \rho$ guarantees the beginning of the radiation dominated epoch.  

Furthermore, from Eqs. \eqref{epsilon1} and \eqref{rhomovN}, one finds the exact relation
\begin{equation}\label{relacionQgen}
	\frac{Q,_N}{Q} = \frac{\left(- 4 \epsilon_1 + 2\epsilon_1^2 - \epsilon_1 \epsilon_2 \right) }{2 - \epsilon_1}  .
 \end{equation}
Consequently, an inflationary phase characterized by $|\epsilon_{1}|, |\epsilon_2| \ll 1$, implies that\footnote{ Equation \eqref{condicionQinf2} is similar to the condition $ (M_P^2/2) ( \partial_\varphi V/V  )^2 \simeq  $ constant $<1$,  which specifies a region in the potential, associated to the inflaton $\varphi$, where slow-roll inflation might occur.}
\begin{equation}\label{condicionQinf2}
	\frac{Q,_N}{Q} = -2 \epsilon_{1} + \mathcal{O}(\epsilon^2) \to 0^-,
\end{equation}
this is, $Q,_N/Q$ is a negative number very close to zero. 

We also note, that during inflation $\epsilon_{1}$ is approximately constant (and very small), so $H \simeq$ constant also during inflation. Therefore,  from Eqs. \eqref{F1} and \eqref{condicionQinf},  one finds that
\begin{equation}\label{Qinftipico}
Q^\diamond  \simeq	3 H^{\diamond 2} M_P^2 \simeq \text{const.},
\end{equation}
where from now on we will denote with a $\diamond$  the typical value of a  dynamical variable during inflation.   

Equation \eqref{Qinftipico}, shows that for any given $Q$ such that inflation takes place, the diffusion term must be approximately constant during inflation, and very close to the characteristic energy scale of inflation.  As is well known, the latter is constrained by the amplitude of the primordial power spectrum (which we will analyze in the next section).  Thus, for a given diffusion term $Q$ with some free parameters, these will have to be adjusted in such a way that $Q$ satisfies Eq.  \eqref{Qinftipico} during inflation.  Furthermore, if the diffusion term continues to evolve,  and its present value is related to the observed value of the CC, then  an extra constraint would have to be imposed on the free parameters of $Q$.

From the initial conditions \eqref{condini},  one notes that $\epsilon^\ini_1 = 1$,
while inflation requires $\epsilon_1 < 1$.  Therefore, our next task is to analyze under which circumstances both conditions are consistent. In other words, we want to explore if the initial condition of $\epsilon_1$  allows inflation to start from the first place. 

Using Eq. \eqref{epsilon1}, we find that
\begin{equation}\label{epsilon1dot}
	\epsilon_{1,N} = - \frac{2}{(1+\Gamma)^2} \Gamma,_N.
\end{equation}
Equation \eqref{epsilon1dot} implies that if $\Gamma^\ini,_N>0 $ then $ \epsilon^\ini_{1,N} <0$, and  inflation can occur from the initial condition $\epsilon^\ini_1 = 1$.   In order to find an explicit expression for $\Gamma^\ini,_N$, we apply $\partial_N$ to the definition $\Gamma \equiv Q/\rho$, obtaining
\begin{equation}\label{Gammadot}
\Gamma,_N = 4 \Gamma + \frac{Q,_N}{\rho} (1+ \Gamma),
\end{equation}
where in the previous equation we have also used the continuity equation  \eqref{rhomovN}.  The initial conditions \eqref{condini}, imply $\rho^\ini = Q^\ini$ so $\Gamma^\ini =1$. Henceforth, from Eq. \eqref{Gammadot}, we obtain
$\Gamma^\ini,_N = 4 + 2 Q^\ini,_N/Q^\ini$.  Thus,  if
\begin{equation}\label{condQinfini}
	\frac{Q^\ini,_N}{Q^\ini} > -2
\end{equation}
then $\Gamma^\ini,_N>0 $, and inflation is guaranteed to start from the chosen initial conditions  \eqref{condini}, which set $\epsilon^\ini_{1} =1$.  

In a similar fashion, one can find the condition required for the diffusion term $Q$ to end the inflationary epoch,  which is characterized by $\epsilon^\fin_{1} \simeq 1$. In this case, the analogous condition to Eq. \eqref{condQinfini} is
\begin{equation}\label{condQinfend}
		\frac{Q^\fin,_N}{Q^\fin} < -2.
\end{equation}

Before continuing with our analysis, it is worthwhile to examine how our previous  findings would be affected if a different initial value $\rho^\ini$ is chosen.  In particular, it may be the case that trans-Planckian censorship arguments would prevent us from selecting an initial energy density close to Planck scale.  If such arguments are applicable, then $\rho^\ini < M_P^4$. 
	
The solution to Eq. \eqref{rhomovN} for a generic  initial value $\rho^\ini$ is
\begin{equation}\label{rhosolalt}
	\rho (N)= \rho^\ini e^{-4 N} - e^{-4 N} \int_{0}^N e^{4 \bar{N}} Q,_{\bar N} d\bar N.
\end{equation}

In addition, from Eqs. \eqref{condicionQinf2} and \eqref{Qinftipico}, one can approximate
\begin{equation}\label{Qprimdi}
	Q^\diamond_{,N} \simeq -6  \epsilon_{1}^\diamond H^{\diamond 2} M_P^2.
\end{equation}
Substituting Eq. \eqref{Qprimdi} into Eq. \eqref{rhosolalt},  and taking into account that $e^{-4 N^\diamond} \ll1$ while inflation is going on $N^\diamond \gg1$, we have 
\begin{equation}\label{rhodi}
	\rho^\diamond \simeq \frac{3}{2} \epsilon_{1}^\diamond  H^{\diamond 2} M_P^2.
\end{equation}
As a consequence, starting from a general initial condition $\rho^\ini$, the energy density during inflation, which is driven by $Q$, has a characteristic scale given by Eq. \eqref{rhodi}.  That result is also consistent with the typical scale of the diffusion term during inflation. That is, Eqs. \eqref{Qinftipico} and \eqref{rhodi},  imply $Q^\diamond/\rho^\diamond \simeq 1/\epsilon_{1}^\diamond > 1$, which is the condition for inflation to take place.  Thus, the equations characterizing $Q$ during inflation, given in Eqs. \eqref{condicionQinf2}  and \eqref{Qinftipico}, are still valid for any initial value $\rho^\ini$; in particular, for a sub-Planckian initial value. 

To complete the analysis regarding a generic $\rho^\ini$, we can assume that $\epsilon_{1}^\ini < \epsilon_1^\diamond < 1$. In particular, that is the case if  one sets the initial value of the energy density equal to the characteristic energy scale of inflation (which is a natural assumption for a sub-Planckian $\rho^\ini$), i.e. $\rho^\ini \simeq H^{\diamond 2} M_P^2$, where $H^{\diamond} < M_P$.  In this situation, $\epsilon_{1}^\ini$  has to increase until it reaches the value $\epsilon_{1}^\diamond$ during inflation, that is, $\epsilon_{1,N}^\ini >0$.  Consequently, from Eq.  \eqref{epsilon1dot}, we observe that $\epsilon_{1,N}^\ini >0$ when $\Gamma^\ini_{1,N} < 0$.  Using  $\rho^\ini =Q^\ini/\Gamma^\ini$ in Eq. \eqref{Gammadot},  one obtains that   if
\begin{equation}\label{condQinfinialt}
	\frac{Q^\ini,_N}{Q^\ini} < - \frac{4}{\Gamma^\ini},
\end{equation}
then $\Gamma^\ini_{1,N} < 0$, where $\Gamma^\ini >1$ for sub-Planckian $\rho^\ini$. In this case,  the condition for $Q^\ini,_N/Q^\ini$ is different from Eq. \eqref{condQinfini}, which was obtained using the initial value   $\rho^\ini \simeq M_P^4$.

In summary, any model based on unimodular gravity, where matter is modeled as a single ultra-relativistic fluid and the diffusion term $Q$ dominates over the energy density, admits an accelerated expansion. Moreover, if $Q$ satisfies Eqs. \eqref{condicionQinf2},  \eqref{condQinfini} and \eqref{condQinfend}, then an inflationary phase that lasts long enough is allowed to start from initial conditions \eqref{condini}.  Also,  $Q$-driven inflation admits a smooth ending, enabling the beginning of the radiation dominated phase. Once that particular diffusion term has been specified, Eq. \eqref{rhosolrad0} yields the evolution of $\rho(N)$,  completing the set of solutions corresponding to the dynamical evolution of the system.  Furthermore, at the background level, the dynamical evolution of the present inflationary model is, for all practical purposes, the same as traditional slow-roll inflation.  We can take the analogy further, and think of $Q$ as playing a similar role as the inflaton potential in slow-roll inflation. 

\subsection{Constructing a diffusion term compatible with inflation}

At this point, the analysis of the background dynamics is complete and the only thing left to specify is $Q$. However,  our strategy to attain a better understanding of the inflationary phase generated by $Q$,  but without choosing a particular micro-physical model, based on e.g. quantum gravity, that results in a specific form of $Q$, is as follows: we will assume a particular function $\epsilon_1$ that is compatible with a full inflationary phase [see discussion after Eq. \eqref{epsilon1}].  For that chosen $\epsilon_{1}$, we construct the corresponding $Q$,  and  consider it as an ansatz.   In this manner, we can analyze some physical consequences for that constructed $Q$. Moreover, if such diffusion term satisfies conditions \eqref{condicionQinf2},  \eqref{condQinfini} and \eqref{condQinfend}, then it can characterize a complete inflationary regime consistent with the initial conditions \eqref{condini}.  

We proceed  to introduce our method for constructing the diffusion term $Q$ from a given $\epsilon_1$.  The dynamical evolution of $\rho$ can be expressed in terms of the HFFs by using: Eq. \eqref{epsilon1}, definition $\Gamma \equiv Q/\rho$,  and $\epsilon_2$. Thus, one can rewrite the continuity equation \eqref{rhomov}  as
\begin{equation}\label{rhomov2}
	\rho(N),_N + [2 \epsilon_1(N) - \epsilon_2 (N) ] \rho(N) =0.
\end{equation}
The later equation can be solved analytically in terms of $\epsilon_1, \epsilon_2$, with the initial conditions given in \eqref{condini}. The exact solution  is
\begin{equation}\label{rhosolNgen}
	\frac{\rho(N)}{M_P^4} = \epsilon_1(N) \exp[-  \int_0^N2 \epsilon_1 (\bar N)   d\bar N  ].
\end{equation}

The solution  \eqref{rhosolNgen} is very useful because it allow us to express $Q(N)$ in terms of the HFFs. That is, by  taking into account that $Q = \Gamma \rho$ and that $\Gamma$ can be obtained by inverting Eq. \eqref{epsilon1}, we have
\begin{equation}\label{QNgen}
	\frac{Q(N)}{M_P^4} = \left[  2 - \epsilon_1(N)    \right]  \exp[-  \int_0^N2 \epsilon_1 (\bar N)   d\bar N  ]. 
\end{equation}
We emphasize that Eqs. \eqref{rhosolNgen} and \eqref{QNgen} are valid for any general form of $\epsilon_1 (N)$.  The only assumption used was that matter behaves as an ultra-relativistic perfect fluid, i.e. $p  = \rho/3$.

Now let us focus on inflation. It can be easily checked that if $|\epsilon_1|, |\epsilon_2| \ll 1$, then Eqs. \eqref{rhosolNgen} and \eqref{QNgen} satisfy conditions \eqref{condicionQinf} and \eqref{condicionQinf2}.   Moreover, for any given $\epsilon_1(N)$ characterizing an inflationary phase, we can construct the corresponding $Q(N)$, via Eq. $\eqref{QNgen}$. In principle,  any $\epsilon_1(N)$ compatible with slow-roll inflation, can be mapped to a function $Q(N)$ through Eq. \eqref{QNgen}. Although, one would need to verify that  the particular diffusion term, constructed from such a slow-roll model, is compatible with conditions \eqref{condicionQinf2}, \eqref{condQinfini} and \eqref{condQinfend}.  The information regarding the beginning and end of inflation is contained in those last conditions.   On the other hand, the solution to the continuity equation $\rho(N)$, for that $\epsilon_1(N)$, is then given in Eq. \eqref{rhosolNgen}.

In order to move forward, we now assume a particular function $\epsilon_1 (N)$ that characterizes a complete inflationary phase.  Therefore, without loss of generality,  we propose 
\begin{equation}\label{epsilon1chingon}
\epsilon_1 (N) = 1 + \tanh \left[ \frac{2}{3} \alpha (N-N_f) \right]	+ \exp(-4 \alpha N),
\end{equation}
where  $\alpha$ is a free parameter of the model\footnote{As seen in Figs. \ref{fig_rhosolN} and \ref{fig_rh}, the free parameter $\alpha$ is linked to the characteristic energy scale of inflation.  };  $N_f$ is the total number of e-folds that inflation lasts;  $N_f$ can be considered another free parameter of the model. Note however, that any other function $\epsilon_1 (N)$ can be used as a first guess in our model, as long as: (a)  it  characterizes an inflationary phase $|\epsilon_1| \ll1 $ that lasts long enough (which means that the condition $|\epsilon_2 | \ll 1$ is required); (b)  it satisfies the initial condition $\epsilon_{1}^\ini=1$; and (c) provides a smooth way to end inflation.
	
The peculiar  function $\epsilon_1(N)$ in \eqref{epsilon1chingon},  serves as an illustrative example of a function satisfying those three conditions. In particular, in order to satisfy the following: (a) $\epsilon_{1} (0)=1$, (b) $\epsilon_1(N) \ll 1$ for $N \ll N_f$ and (c) $\epsilon_1(N_f) = 1$,  the proposed $\epsilon_{1} (N)$  requires a combination of the free parameters such that $\alpha N_f \gg 1$.   The motivation for choosing that function as an example is because it can also accommodate the late accelerated expansion (as will be shown in Sec. \eqref{sec_CC}).  Specifically, the diffusion term, corresponding to that particular $\epsilon_{1} (N)$, approximately reproduces  the observed order of magnitude of the present CC and other cosmological quantities of interest by choosing some particular values of the free parameters: $\alpha$ and $N_f$.  Certainly, much simpler functions,  compatible with the aforementioned conditions, can be considered if one is only interested in characterizing an inflationary phase. We have checked, for example, that the functions: $\epsilon_{1} (N) = (N-N_f/2)^{2n}/(N_f/2)^{2n}  $ or $\epsilon_{1} (N) = \sinh^{2n}[\alpha(N-N_f/2)]/\sinh^{2n}[\alpha N_f/2]$, with $n$ an integer $\geq 1 $,  also satisfy the three main conditions\footnote{  Note that both functions are normalized such that $\epsilon_{1}^\ini = \epsilon^\fin_{1} = 1$. Moreover,  $\epsilon_{1} (N) \ll 1$ around $N = N_f/2$.} (a), (b) and (c). In addition, those two functions result in a diffusion term $Q$ consistent with Eqs. \eqref{condicionQinf2}, \eqref{condQinfini} and \eqref{condQinfend}. In other words, that pair of functions can also reproduce a realistic inflationary model, with the initial conditions considered here.

\begin{figure}[h]
	\begin{subfigure}{.5\textwidth}
		\centering
		\includegraphics[width=1.0\linewidth]{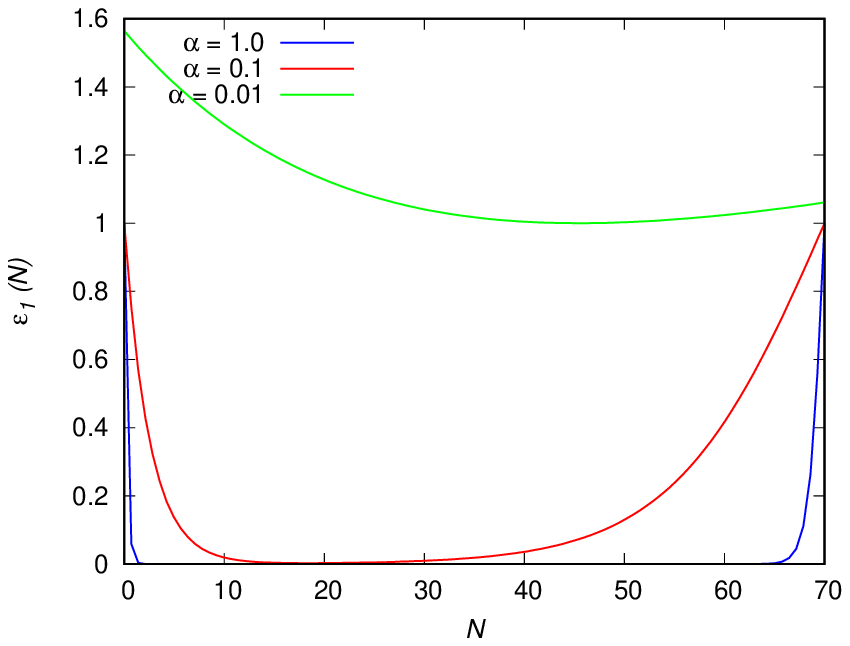}
	\end{subfigure}%
	\begin{subfigure}{.5\textwidth}
		\centering
		\includegraphics[width=1.0\linewidth]{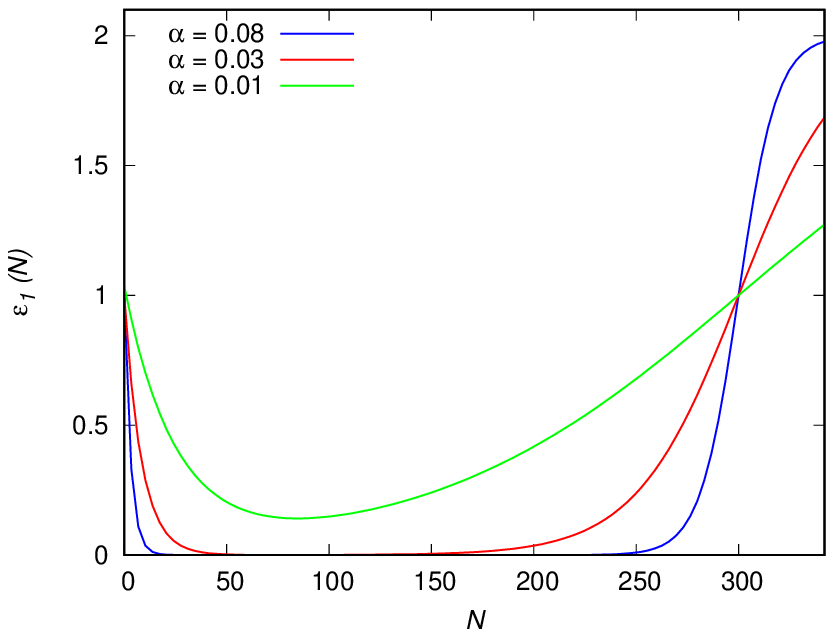}
	\end{subfigure}
	\caption{The function $\epsilon_1 (N)$ as introduced in Eq. \eqref{epsilon1chingon} for different values of $\alpha$. Left:  The total duration of inflation assumed is $N_f = 70$, which is a typical value used for inflation. In this case, the  chosen values of $\alpha$ serve to provide a visualization of the effects in varying such a parameter during inflation. Right: The total duration of inflation assumed is $N_f = 300$, and  the  chosen values of $\alpha$ are very close to $\alpha = 0.03$. In this case,  those particular values of the free parameters will be of interest when analyzing the post-inflationary era.}
	\label{fig_epsilonN}
\end{figure}

Figure \ref{fig_epsilonN} depicts $\epsilon_1 (N)$,  as shown in Eq. \eqref{epsilon1chingon},  for various values of $\alpha$. In Fig. \ref{fig_epsilonN}-left  we have used $N_f = 70$.  Recall that a minimum of $N_f \sim$ 60-70 is required for solving the horizon and flatness problems of the hot Big-Bang model \cite{BaumannTASI2009}, but in fact there is no upper bound on $N_f$.  We observe that lower values of $\alpha$, i.e. $\alpha \leq 0.01$, fail to reproduce an accelerated expansion at all.  In contrast, for   $\alpha \geq 0.1$,  the function $\epsilon_{1} (N)$ correctly describes an inflationary phase.  In Fig. \ref{fig_epsilonN}-right, we have used $N_f = 300$ and chosen values of $\alpha$ very close to $\alpha = 0.03$.  In particular, for $\alpha = 0.03$ and $\alpha = 0.08$, $\epsilon_{1}$ serves to characterize a consistent inflationary epoch. The motivation for selecting those particular set of values  for the free parameters is because, as we will shown in Sec. \ref{sec_CC}, they correctly reproduce the  order of magnitude of the currently observed CC. However,  if one is focused only in reproducing an inflationary phase, $N_f$ and $\alpha$ could be constrained by using the data associated to the cosmological parameters during inflation (namely, the ones coming from the primordial power spectrum).  This is analogous to what is usually done in the traditional  slow-roll  inflationary models, where a specific form of the inflaton potential is considered, and  the corresponding parameters are constrained using the observational data through the predicted primordial power spectrum \cite{jmartinpotentials}.


Given the proposed function  $\epsilon_1 (N)$  in Eq. \eqref{epsilon1chingon}, the diffusion term $Q(N)$ obtained from \eqref{QNgen} is
\begin{eqnarray}\label{Qsol}
	\frac{Q(N)}{M_P^4} &=&   \left\{ 1 - \tanh \left[ \frac{2}{3} \alpha (N-N_f) \right]	- \exp(-4 \alpha N)  \right\}  \nn
	&\times& \exp \left[ \frac{1}{2 \alpha}  \left(  -1 + e^{-4 \alpha N} -4 \alpha N \right)   \right] \nn
	&\times& \left\{    \frac{\cosh (2\alpha N_f/3)}{\cosh \left[ 2\alpha \left(N - N_f \right)/3 \right]}\right\}^{3/\alpha}.
\end{eqnarray}

\begin{figure}
	\begin{subfigure}{.5\textwidth}
		\centering
		\includegraphics[width=1.0\linewidth]{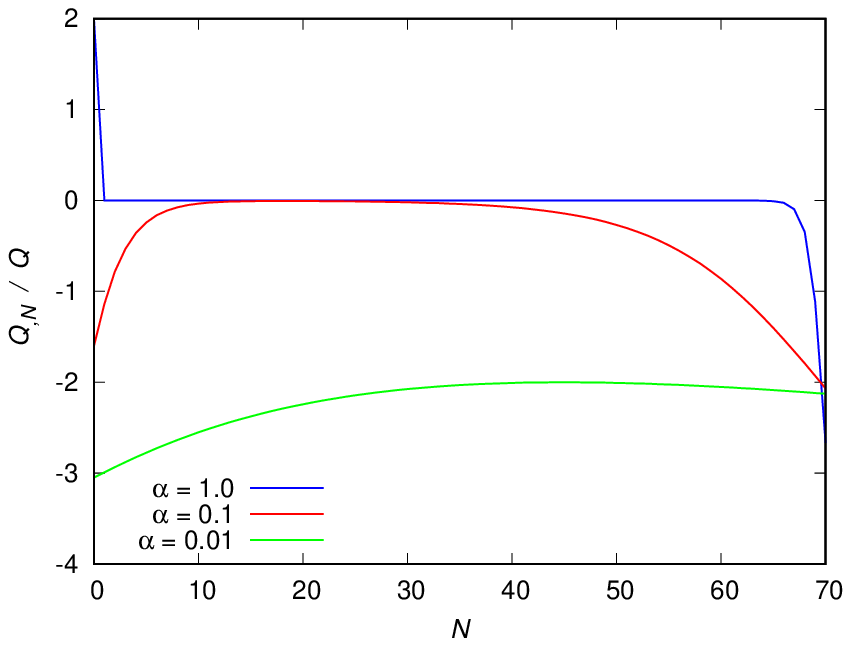}
	\end{subfigure}%
	\begin{subfigure}{.5\textwidth}
		\centering
		\includegraphics[width=1.0\linewidth]{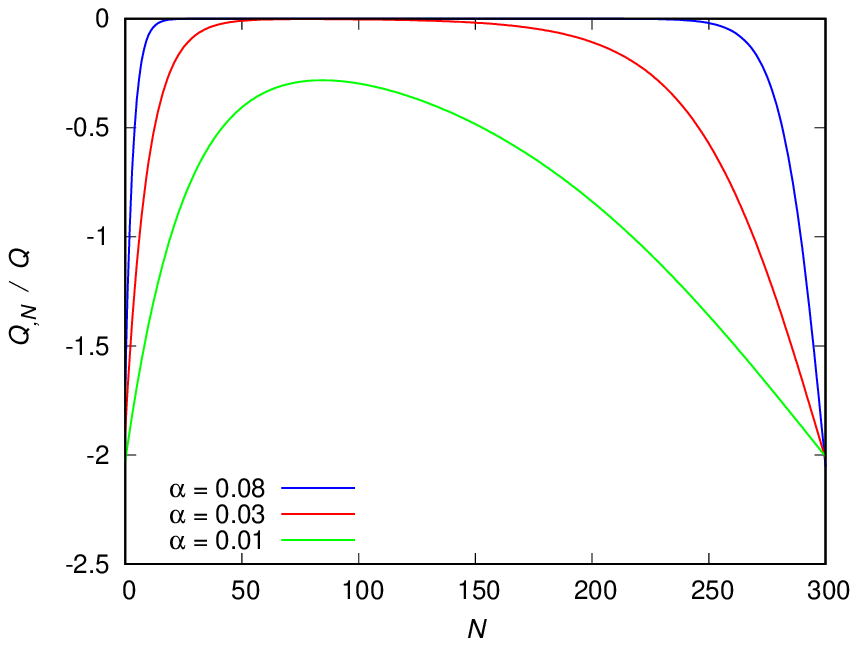}
	\end{subfigure}
\caption{The function $Q,_N/Q$ obtained from the diffusion term  \eqref{Qsol} for different values of $\alpha$. Left:  The total duration of inflation assumed is $N_f = 70$. In this case, the  chosen values of $\alpha$ serve to provide a visualization of the effects in varying such a parameter during inflation. Right: The total duration of inflation assumed is $N_f = 300$, and  the  chosen values of $\alpha$ are very close to $\alpha = 0.03$. In this case,  those particular values of the free parameters will be of interest when analyzing the post-inflationary era.   }
\label{fig_R}
\end{figure}

\begin{figure}
  	\begin{subfigure}{.5\textwidth}
  		\centering
  		\includegraphics[width=1.0\linewidth]{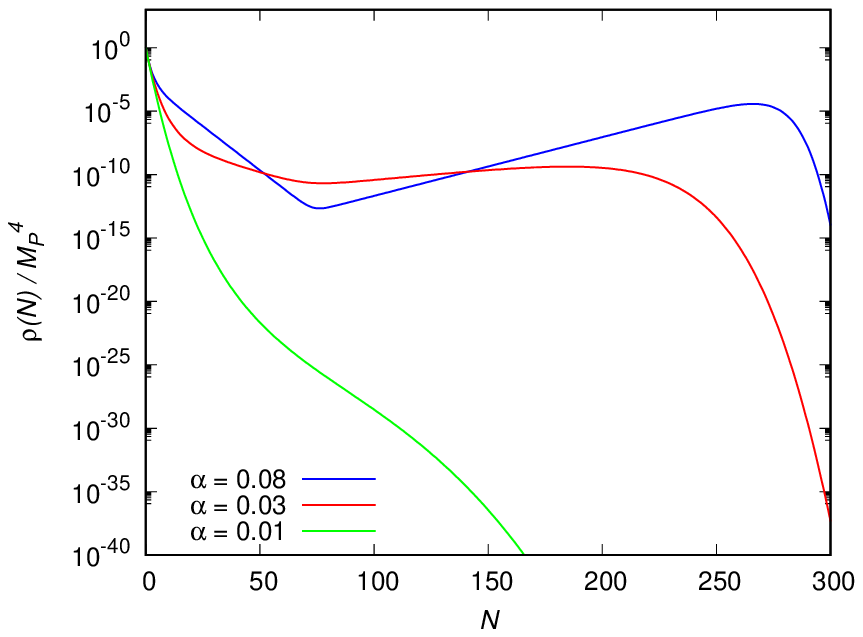}
  	\end{subfigure}%
  	\begin{subfigure}{.5\textwidth}
  		\centering
  		\includegraphics[width=1.0\linewidth]{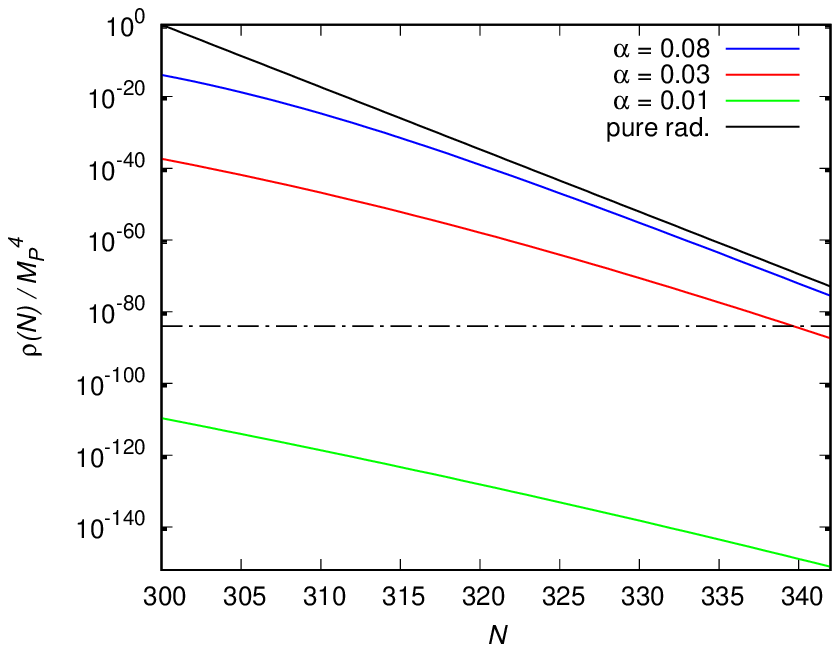}
  	\end{subfigure}
  	\caption{ The solution $\rho (N)$ obtained in Eq. \eqref{rhosol} for different values of $\alpha$ close to $\alpha=0.03$.  Left: The evolution of $\rho(N)$ during inflation assuming a total duration of $N_f = 300$.  Right: The evolution of $\rho(N)$ in the post-inflationary epoch. The constant dashed line marks the scale $\rho = 10^{-84} M_P^4$, which corresponds to an energy scale of $100$ GeV, i.e. the electroweak symmetry breaking energy scale.   The black line labeled as ``pure rad.'' represents the function $e^{-4(N-N_f)}$, which is the usual evolution of $\rho$ if the fluid behaves as  pure radiation and no violation of the continuity equation occurs. }
  	\label{fig_rhosolN}
  \end{figure}

\begin{figure}
	\centering
	\includegraphics[scale=1.0]{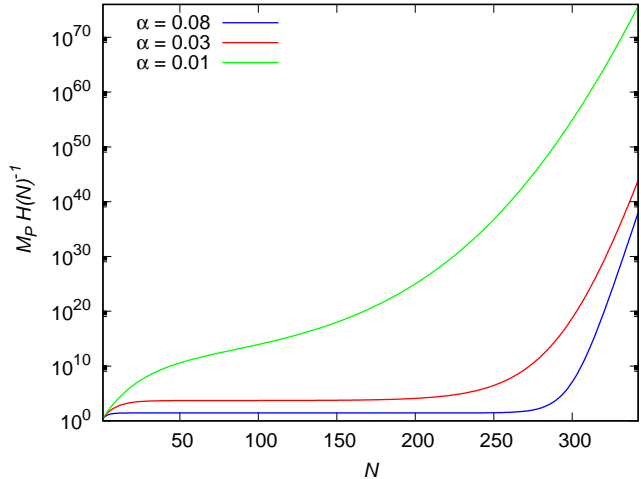}
	\caption{ The Hubble radius $H^{-1} (N)$, obtained from \eqref{F1}, using $\rho(N)$ and $Q(N)$ from previous figs. We have used  three different values of $\alpha$ and assuming a total duration of inflation of $N_f = 300$. }
	\label{fig_rh}
\end{figure}

With the diffusion term at hand, we can evaluate whether $Q(N)$  satisfies the conditions given in Eqs. \eqref{condicionQinf2}, \eqref{condQinfini} and \eqref{condQinfend}. Recall that those equations characterize the inflationary model consistent  with the initial conditions  \eqref{condini}, and which eventually ends in a smooth manner. In Fig. \ref{fig_R} various plots of $Q,_N/Q$ are shown.  In Fig. \ref{fig_R}-left, where $N_f =70$ was used, we note that for $\alpha \geq 0.1$ the function $Q,_N/Q$ satisfies the aforementioned equations.  In contrast, for $\alpha \leq 0.01$ the opposite occurs. This is compatible with the previous discussion involving the behavior of $\epsilon_{1}$ for that same set of parameters.  In Fig. \ref{fig_R}-right, where $N_f =300$, we note that for  $\alpha=0.03$  and $\alpha = 0.08$, the function  $Q,_N/Q$ satisfies the conditions given in Eqs. \eqref{condicionQinf2}, \eqref{condQinfini} and \eqref{condQinfend}. Thus, a consistent inflationary phase is allowed for the constructed diffusion term and those values for $N_f$ and $\alpha$.  From now on,  and because of the motivation exposed above, we will focus on those  particular  values:  $N_f=300$ and $\alpha \simeq 0.03$.


In addition,  from Eq. \eqref{rhosolNgen}, the energy density is given as
\begin{eqnarray}\label{rhosol}
	\frac{\rho(N)}{M_P^4} &=& 	 \left\{ 1 + \tanh \left[ \frac{2}{3} \alpha (N-N_f) \right]	+ \exp(-4 \alpha N)  \right\}  \nn
	&\times& \exp \left[ \frac{1}{2 \alpha}  \left(  -1 + e^{-4 \alpha N} -4 \alpha N \right)   \right] \nn
	&\times& \left\{    \frac{\cosh (2\alpha N_f/3)}{\cosh \left[ 2\alpha \left(N - N_f \right)/3 \right]}\right\}^{3/\alpha}.
\end{eqnarray}
One can check, that $Q(N)$ and $\rho(N)$, Eqs. \eqref{Qsol} and \eqref{rhosol}, satisfy the continuity equation \eqref{rhomovN}.  In Fig. \ref{fig_rhosolN}, we present the dynamical evolution of the exotic looking function $\rho(N)$, Eq. \eqref{rhosol}.

  In  Fig. \ref{fig_rhosolN}-left,  the evolution of $\rho$ is shown during the full inflationary period for $N_f =300$ and three values close to $\alpha = 0.03$. As we observe there,  the energy density begins decreasing with the expansion. For $\alpha \leq 0.01$, $\rho$ continues to decrease indefinitely.  For $\alpha = 0.03$ and $\alpha =0.08$, $\rho$ decreases until a minimum is reached, in this instant the effect of $Q$ ``kicks in'' stopping the rapid decrease of $\rho$.  In particular, for $\alpha = 0.08$, $\rho$  increases during some interval within the inflationary phase, and then decreases again near the end of inflation.   It is worth mentioning that in this last case, the dynamical evolution of $\rho(N)$, as shown in Fig. \ref{fig_rhosolN}-left for $\alpha =0.08$, is very similar to that obtained in Ref. \cite{aperez2021} (see Fig. 1 in that reference, and take into account that inflation lasts $\sim$60 e-folds in there), in which the origin of the diffusion term $Q$ is attributed to a possible fundamental discreteness of the spacetime at such high energy scales. 
  
In Fig.   \ref{fig_rhosolN}-right, the evolution of $\rho$ is shown after inflation has ended.  For all values of $\alpha$, the energy density decreases similar as $\rho \propto a^{-4}$ during the subsequent radiation dominated epoch.  In fact, as we can see there,  for $\alpha = 0.03$ and   $N \simeq 340$, the energy density is $\rho/M_P^4 \simeq 10^{-84}$, which is equivalent to an energy scale of $100$ GeV i.e. the electroweak symmetry breaking energy scale. Additionally, from Fig \ref{fig_epsilonN}-right and for the same value $\alpha = 0.03$, we observe that at  $N \simeq 340$ the inflationary phase has ended long ago. Therefore, the dynamical evolution between $\epsilon(N)$ and $\rho(N)$ is compatible.



Having found $Q(N)$ and $\rho(N)$, it is straightforward to obtain $H(N)$ from Friedmann's equation \eqref{F1}.  In Fig. \ref{fig_rh} we have plotted $H^{-1} (N)$,  i.e. the Hubble radius.  We observe that the Hubble radius remains constant during the whole inflationary regime, then after reaching its end,  $H^{-1} (N)$ starts to increase as expected. Moreover,  we note that for low values of $\alpha$, there is an increase in  the magnitude of  $H^{-1}$.  In this way, we can see that the free parameter $\alpha$ is directly related with the absolute magnitude of $H$ during inflation.

The complete behavior of $Q(N)$ will be analyzed  in Sec. \ref{sec_CC}, but from the results obtained in the present section,  we can conclude that ansatz \eqref{Qsol} is compatible with an early inflationary phase.  It also provides the right amount of inflation from natural initial conditions and induces a graceful ending.  The next subject we want to focus on is the primordial power spectrum generated during the inflationary phase.

\section{Cosmological perturbations in unimodular gravity and  the primordial power spectrum}\label{sec_PS}

In this section we will analyze the dynamics of the cosmological perturbations, with the main purpose  of obtaining the primordial power spectrum.  In fact,  throughout this section, we will assume that the primordial spectrum is generated by considering inhomogeneous density perturbations $\delta \rho$ associated to the background energy density $\rho_0$ of the ultra-relativistic fluid (analyzed in the previous section).

The analysis, however, involves the use of cosmological perturbation theory in UG. This topic has been explored previously in the literature and some subtle differences from perturbation theory in GR arise. We will follow Refs. \cite{UGindios,UGbranden} for some general aspects involving this subject.  Furthermore, the energy-momentum tensor  non-conservation, which plays a key role in our model, introduces additional caveats when adopting the UG version of cosmological perturbation theory. Therefore, we will proceed in a somewhat slow manner in order to be careful with the assumptions used when deriving the main equations in the present section.

We will focus only in the scalar sector of the metric-matter perturbations.  Switching to conformal time $\eta$ (i.e. $d \eta = dt/a$), the metric components up to first order in the perturbations  are given generically by
\begin{subequations}\label{metricaperts}
	\begin{equation}\label{eqg00}
			g_{00} = - a^2(1+2 \phi),
	\end{equation}
\begin{equation}\label{eq0i}
	g_{0i} = a^2  \partial_i B,
\end{equation}
\begin{equation}\label{eqij}
	g_{ij} = a^2 (\delta_{ij} - 2 \psi \delta_{ij} + \partial_{ij} 2 E  ),
\end{equation}
\end{subequations}
where $\phi$, $\psi$, $B$  and $E$ are scalar functions dependent on the time and spatial coordinates.    

The metric $g_{\mu \nu }$ can thus be expressed as a background part plus a perturbation, i.e.  $g_{\mu \nu} = a^2 \eta_{\mu \nu} + \delta g_{\mu \nu} $, where  $\eta_{\mu \nu}$ is  the Minkowski metric and $\delta g_{\mu \nu} $ is the perturbation representing a departure from perfect spatial homogeneity/isotropy. The unimodular constraint $\delta \sqrt{-g} = 0$ yields the following constraint equation for the scalar perturbations, 
\begin{equation}\label{ugconstraintpert}
	\phi -3 \psi + \nabla^2 E =0.
\end{equation}

Equation \eqref{ugconstraintpert}, reduces by one the freedom of choice available that is contained in the election of gauge normally present in standard treatments of cosmological perturbation theory.  This is one of the main differences between UG and GR in the present context.  In particular, the so called  \textit{longitudinal} gauge (where $B=E=0$) or \textit{synchronous} gauge (where $\phi = B = 0$) cannot be considered in UG due to constraint \eqref{ugconstraintpert} \cite{UGindios,UGbranden}.

On the other hand, gauge-invariant quantities can be constructed in UG.  Let us note however that, as explained in e.g. \cite{Malik2009}, \textit{gauge-invariant} is not the same as \textit{gauge independence.}   For example, the  intrinsic spatial curvature on constant time hypesurfaces, which is proportional to $\psi$, is not gauge independent.  It is different under different time slicings.  In other words, one can construct gauge-invariant combinations, which in the literature are usually referred to  as  the gauge-invariant curvature perturbation, but they only correspond with the curvature perturbation in one particular gauge.  According to this line of reasoning, we will work with a gauge invariant quantity in UG and then identify it with the curvature perturbation in a particular gauge that is compatible with the unimodular constraint \eqref{ugconstraintpert}.  Afterwards, we will analyze its dynamical behavior so we can finally obtain the primordial power spectrum. 
%

The EFE in UG can be written as (see Eq. \eqref{EE1})
\begin{equation}\label{EE2}
		R^{\mu}_{\nu} - \frac{1}{2} \delta^{\mu}_{\nu}  R + \lambda(x) \delta^{\mu}_{\nu} = \mathcal{T}^{\mu}_{\nu}.
\end{equation}
We recall that from the constraint equation $\lambda(x)=(R+\mathcal{T})/4$, and after fixing the value of the integration constant $\Lambda_*=0$,  we obtained $\lambda(x) = Q(x)/M_P^2$.  Therefore, we have the relation $(R+\mathcal{T})/4= Q/M_P^2$. Furthermore, by taking into account the EOS $p = \rho/3$, valid for pure radiation, one obtains $ \mathcal{T} =0$. As a consequence, from the latter relation, we have 
\begin{equation}\label{eqRyQ}
	R = \frac{4 Q}{M_P^2}.
\end{equation}

In Secs. \ref{seccosmoUG} and \ref{sec_inf}, we have assumed that the diffusion term $Q$ is completely homogeneous. However, strictly speaking, that might not be the case in a more general setting.  In particular, it could be the case that the micro-physics leading to a specific form of $Q$ introduces small inhomogeneities (for example, in Refs. \cite{sudarskyPRL1,Corral20} the origin of the diffusion term is linked to the  matter fields, characterized by the energy-momentum tensor).  While at the background level, $g_{\mu \nu}$, $T_{\mu \nu}$ and $Q$ can all be considered homogeneous according to the cosmological principle, when analyzing the primordial inhomogeneities, some residual inhomogeneity $\delta Q(x)$ could be present. Here, the relevant question is whether $\delta Q$ is sourced by  first or second-order terms in the perturbations associated to the metric and/or energy density.  Clearly, a precise answer to such a question would require a particular physical model explaining the origin of the diffusion term, this is beyond the scope of the present paper (but is a subject worth exploring in the future).  Therefore, as a first guess, we will assume that if some residual inhomogeneities are present in $Q$, they can be considered at second order in the perturbations.\footnote{ This is also motivated by the fact that we are interested in analyzing if a primordial power spectrum, with the correct shape and amplitude, can be found from an inflationary phase driven by $Q$. Hence,  to take a first step, we restrict ourselves to the simplest case.  } Thus, for all practical purposes, in the rest of this section we assume $Q$ to be homogeneous.

Given that $Q(\eta)$ is a homogeneous quantity, one has $\delta Q =0$; consequently,  from Eq. \eqref{eqRyQ},  we obtain $\delta R = 0$.  Using these results, we find from Eqs. \eqref{EE2}, the following set of equations for the perturbations
\begin{equation}\label{EEpert}
		\delta R^{\mu}_{\nu} - \frac{1}{2} \delta^\mu_\nu  \delta R  = \delta \mathcal{T}^{\mu}_{\nu} ,
\end{equation}
with the understanding that $\delta R = 0$. The reason to include the vanishing term $\delta R = 0$ in \eqref{EEpert}, is because it exhibits explicitly that working under the assumption of a perfect fluid consisting of pure radiation, the perturbed EFE at linear order in UG, Eqs. \eqref{EEpert}, are exactly the same as in traditional GR. This can also be seen by putting the EFE of GR in trace form, i.e. $-R +4\Lambda = \mT$, where $\Lambda$ is a true (cosmological) constant, and since $\mT=0$ for a radiation fluid, one finds that $\delta R =0$. Consequently, the EFE in GR lead to exactly the same set of equations  as the ones in \eqref{EEpert}.  We stress that this particular  result is only valid because of two key assumptions: (i)  the total matter content in the universe behaves (during the early phase(s) considered here) as a perfect fluid consisting of pure radiation, and (ii) the quantity $Q$, characterizing the conservation violation of the energy-momentum tensor, is completely homogeneous.

In view of discussion above, we can find the set of equations \eqref{EEpert} using the metric perturbations given in \eqref{metricaperts}, which are written in an arbitrary gauge.  For that, we introduce first the components of the perturbed energy-momentum tensor:
\begin{subequations}\label{Tabperturb}
	\begin{equation}\label{dT00}
		\delta T^0_0 = -\delta \rho
	\end{equation}
\begin{equation}\label{dT0i}
	\delta T^0_i = (\rho_0 + p_0 ) (v+B),_i 
\end{equation}
\begin{equation}\label{dTi0}
\delta T^i_0 = -(\rho_0 + p_0) v,^i
\end{equation}
\begin{equation}\label{dTij}
	\delta T^i_j = \delta p \delta^i_{j}
\end{equation}
\end{subequations}
where $\delta \rho$ and $\delta p$ represent the first-order scalar perturbations of the energy density and pressure relative to their background counterparts denoted as $\rho_0$, $p_0$.\footnote{  Throughout this section, we are assuming that the total matter content in the universe is modeled by an ultra-relativistic fluid. That is, the EOS characterizing the fluid is $p =  \rho/3$.  Therefore, the relation between the pressure and energy density perturbations is given as  $\delta p = \delta \rho/3$. In other words, the adiabatic sound speed in this case is $c_s^2 = 1/3$. }  The object $v$ represents the scalar part of the spatial velocity of the fluid, sometimes it is referred as  the \textit{velocity potential} \cite{mukhanov92,Wald2002} and is a first-order perturbation.  In addition, we can see that there is no anisotropic stress.

After having introduced the components of $\delta T^\mu_\nu$, we can write the full set of equations \eqref{EEpert}. Thus, the full set of EFE at linear order in the perturbations is:
\begin{subequations}\label{dEFE}
	\begin{equation}\label{EFE00}
		3\mH (\mH \phi + \psi') - \nabla^2 (\psi + \mH \sigma) = - 4 \pi G a^2 \delta \rho,
	\end{equation}
\begin{equation}\label{EFE0i}
	\mH \phi + \psi' = -4 \pi G a^2 (\rho_0 + p_0) (v+B),
\end{equation}
\begin{equation}\label{EFEij}
	\sigma' + 2\mH \sigma + \psi - \phi = 0,
\end{equation}
\begin{equation}\label{EFEii}
	\psi'' + 2\mH \psi' + \mH \phi' + ( 2\mH' + \mH^2) \phi = 4 \pi G a^2 \delta p,
\end{equation}
\end{subequations}
where a prime denotes derivative with respect to $\eta$, and we have introduced the definitions $\mH \equiv a'/a$, $\sigma \equiv E'-B$.  Note that Eqs. \eqref{dEFE}, are expressed in an arbitrary gauge. 

Another useful set of equations correspond to the continuity equation $\nabla_\mu T^\mu_\nu = \delta^\mu_ \nu   \nabla_\mu Q$. As we have argued, assumption (ii), leads to
\begin{subequations}\label{eqsmovpert}
	\begin{equation}\label{consback}
			\nabla_\mu \:^{(0)} T^\mu_\nu  =  \delta^\mu_ \nu   \nabla_\mu Q,
	\end{equation}
\begin{equation}\label{conspert0}
	\nabla_\mu \delta T^\mu_\nu =0,
\end{equation}
\end{subequations}
where $^{(0)} T^\mu_\nu $ denotes the background part of the energy-momentum tensor. In particular, Eq. \eqref{consback} corresponds to Eq. \eqref{rhomov}.  We observe that the diffusion  term $Q$ is coupled only to the background part. Nonetheless, this does not imply that $Q$ has no effect on the conservation equation for the perturbations. Indeed, Eq. \eqref{conspert0}, depends on dynamical background quantities, such as $\rho_0$, $\mH$, etc., that as we have shown in the previous section, are strongly influenced by the source term $Q$.

From the components of $ \delta T^\mu_\nu$,  i.e.  Eqs. \eqref{Tabperturb} and the perturbed part of the metric, one can obtain the corresponding conservation equations for the perturbations characterized by Eq. \eqref{conspert0}. These are  
\begin{subequations}\label{conspert}
	\begin{equation}\label{drhoprim}
		\delta \rho' + 3\mH (\delta \rho + \delta p ) + (\rho_0 + p_0) [\nabla^2 (\sigma + v + B)-3 \psi'] =0,
	\end{equation}
\begin{equation}\label{vprim}
	[(\rho_0 + p_0)(v+B)]' + (\rho_0 + p_0) [4 \mH (v + B) + \phi] + \delta p =0.
\end{equation}
\end{subequations}
 We note again that Eqs. \eqref{conspert} are expressed in a generic gauge and no dynamical background equations have been used.

%
 
 A gauge transformation can be considered technically equivalent to an infinitesimal coordinate transformation. At first order, the latter is given as $x^\mu \to \tilde x^\mu = x^\mu + \xi_1^\mu$, where the gauge generator $\xi_1^\mu$ has components $\xi_1^\mu = \{  \alpha_1 , (\beta_{1,i})  \}$.  Under that coordinate transformation, the first-order scalar metric perturbations transform as \cite{Malik2009,UGbranden}
 \begin{subequations}\label{transmetric0}
 	\begin{equation}\label{transphipsi0}
 		\tilde \phi = \phi +\mH \alpha_1+ \alpha_1', \qquad \tilde \psi = \psi - \mH \alpha_1,
 	\end{equation}
 	\begin{equation}\label{transEB0}
 		\tilde B = B - \alpha_1 + \beta_1', \qquad \tilde E = E + \beta_1.
 	\end{equation}
 \end{subequations}
The scalar perturbation $\delta \rho$ and the combination $\chi \equiv v+B$, associated to the energy-momentum tensor components, transform as
 \begin{subequations}
 	\begin{equation}\label{transdrho}
 		\tilde \delta \rho = \delta \rho + \rho_0'  \alpha_1,
 	\end{equation}
 	\begin{equation}\label{transvB}
 	\tilde \chi = \chi - \alpha_1.
 	\end{equation}
 \end{subequations}
 
 By substituting the transformed metric perturbations Eqs. \eqref{transmetric0} into the constraint equation \eqref{ugconstraintpert}, one can rewrite the unimodular constraint in terms of the  quantities $\alpha_1$ and $\beta_1$; this is
 \begin{equation}\label{ugconstraintpert0}
 	\nabla^2 \beta_1 + \alpha_1' + 4 \mH \alpha_1 =0.
 \end{equation}
 
Next, we choose a  gauge such that $\tilde \chi \equiv \tilde v + \tilde{B} =0$. From Eq. \eqref{transvB}, we find that in this gauge 
 \begin{equation}\label{alpha1}
 	\alpha_1 = v+B \equiv \chi. 
 \end{equation}
Furthermore, from the unimodular constraint \eqref{ugconstraintpert0} and Eq. \eqref{alpha1},  the function $\beta_1$ becomes fixed by the following equation
 \begin{equation}\label{ugbetaconstraint}
 	\nabla^2 \beta_1  + \chi' + 4 \mH \chi =0.
 \end{equation}
Thus, for the selected gauge (in which $\tilde \chi =0$),  Eqs. \eqref{alpha1} and \eqref{ugbetaconstraint} completely specify the components of the gauge generator, i.e. $\alpha_1$ and $\beta_1$. 
 
Additionally, the former chosen gauge induces a gauge-invariant quantity defined as
 \begin{equation}\label{Rcomov}
 	\mR \equiv \psi - \mH \chi,
 \end{equation}
Using the generic transformations given in Eqs.  \eqref{transphipsi0} and \eqref{transvB}, it is straightforward to check that $\mR$ is gauge-invariant. Finally, working in the chosen gauge $\tilde \chi =0$,  one has that $\mR = \tilde \psi$.  In other words, $\mR$ is the so called  \textit{comoving curvature perturbation}. In fact, the gauge we have chosen is the well known comoving gauge, but expressed in terms of the ``velocity potential''  rather than the first order perturbation of the inflaton as done in standard treatments of slow-roll inflation.

%

Our next task is to find the evolution equation for the comoving curvature perturbation $\mR$ and obtain the corresponding power spectrum. In this way,  we can analyze if  our  model is able to reproduce (or not) the  primordial power spectrum with the known features that are consistent with the data, i.e. with the right amplitude and shape.

In the comoving gauge, where $v+B=0$ and $\psi = \mR$, Eqs. \eqref{dEFE} together with $\delta p = \delta \rho/3$ can be combined into the following expression
 \begin{equation}\label{Rmov0}
	\mR'' + 2\mH \mR' + \mH \phi' +  (2\mH' + \mH^2) \phi = \frac{1}{3} \nabla^2 (\mR + \mH \sigma).
\end{equation}
Equation \eqref{EFE0i} provides the useful relation, 
\begin{equation}\label{relphiyR}
	-\mR' = \mH \phi.
\end{equation}
Consequently, in order to obtain a closed form for the evolution equation of $\mR$, we have to find an adequate expression to replace the term $(1/3) \nabla^2 \mH \sigma$ on the right hand side of Eq. \eqref{Rmov0}. The sought expression comes from the continuity equations \eqref{conspert}. 
Specifically, the conservation equation Eq. \eqref{vprim} in the comoving gauge becomes
\begin{equation}\label{deltap}
	\delta p = -\phi (\rho_0 + p_0).
\end{equation}
The other conservation equation \eqref{drhoprim}, which involves $\delta \rho'$,  combined with Eq. \eqref{deltap}, and using that $\delta p = \delta \rho/3$, yields 
\begin{equation}\label{nablasigma}
	\nabla^2 \sigma = 3 \phi' + 12 \mH \phi +3 \mR' + 3 \phi \frac{(\rho_0 + p_0)'}{\rho_0 + p_0}.
\end{equation}
Substituting Eq. \eqref{nablasigma} on the right hand side of Eq. \eqref{Rmov0} and using relation \eqref{relphiyR}, we obtain the evolution equation of $\mR$  in closed form  
\begin{equation}\label{Rmov1}
	\mR'' = \frac{1}{3} \nabla^2 \mR + \mR' \left(   - \frac{(\rho_0 + p_0)'}{\rho_0 + p_0} + 2 \frac{\mH'}{\mH} - 4\mH        \right).
\end{equation}

Equation \eqref{Rmov1} is the sought equation. However, we can take a further step to convert it into a more recognizable form. Using Friedmann's equations \eqref{F1} and \eqref{F2}, in conformal time coordinate,  one obtains
\begin{equation}\label{backchida}
	\mH^2 - \mH' = \frac{a^2}{2 M_P^2} (\rho_0 + p_0).
\end{equation}
The main feature of Eq. \eqref{backchida} is that it is independent of  $Q$ explicitly. Hence,
Eq. \eqref{Rmov1}, becomes
\begin{equation}\label{Rmovchida}
	\mR'' - \frac{1}{3} \nabla^2 \mR + 2 \frac{z'}{z} \mR,
\end{equation}
where we have introduced the quantity $z$ defined as
\begin{equation}\label{defz}
	z \equiv \sqrt{2} M_P \frac{a}{\mH} (\mH^2 - \mH')^{1/2}.
\end{equation}
Finally, introducing one more definition $\mathcal{M} \equiv  z \mR$, we obtain the familiar  Mukhanov-Sasaki (MS) equation
\begin{equation}\label{MSmov}
	\mM'' - \left(  \frac{1}{3} \nabla^2  + \frac{z''}{z} \right) \mM = 0.
\end{equation}

The evolution equation of $\mR$, Eq. \eqref{Rmovchida}, (or equivalently Eq. \eqref{MSmov} for $\mM$) is exactly the same as the one obtained in traditional models of slow-roll inflation \cite{mukhanov92,ringevalSR,Garriga1999}. The main difference is of course that no scalar field has been considered in our model.  We have taken a long route to find Eqs. \eqref{Rmovchida}, \eqref{MSmov} in order  to show that the non-conservation equation for the background part of the energy-momentum tensor does not enter explicitly into the derivation.  This is an important aspect since in treatments of cosmological perturbation theory in UG, the conservation of the energy-momentum tensor, which is imposed as an extra assumption, is regularly used to derive the equations of motion associated to gauge invariant quantities, in particular, the ones associated to the curvature perturbation.\footnote{See, for instance, the discussion around Eq. (7) and the derivation of Eq. (28) in Ref. \cite{UGbranden}.  Additionally, see also Sec. 2.3 and the derivation of Eq. (4.29) in Ref. \cite{UGindios}. } However,  we note that assumptions (i) and (ii) were key in obtaining  Eqs. \eqref{Rmovchida}, \eqref{MSmov}.

It is then straightforward to obtain the power spectrum since it involves the same steps as in traditional slow-roll inflation.  We can summarize them as follows. Taking into account that $z$ can be written as $z = a M_P \sqrt{2 \epsilon_1}$, and switching to Fourier space, the MS equation becomes
\begin{equation}\label{eqMSfourier}
		\mM_k'' + \left(  \frac{k^2 }{3}  - \frac{(a\sqrt{\epsilon_1})''}{a \sqrt{\epsilon_1}} \right) \mM_k = 0.
\end{equation}

Furthermore, since $\epsilon_1$ and $\mH$ are constructed to possess the known features associated to an inflationary phase (see e.g. Figs. \ref{fig_epsilonN} and \ref{fig_rh}) 
we can use  the standard result \cite{ringevalSR,venninSR}
\begin{equation}\label{appz}
	\frac{(a\sqrt{\epsilon_1})''}{a \sqrt{\epsilon_1}} = \mH^2 \left[2 - \epsilon_1 + \frac{3}{2}   \epsilon_2 +\frac{1}{4} \epsilon_2^2 - \frac{1}{2} \epsilon_1 \epsilon_2     + \frac{1}{2}   \epsilon_2 \epsilon_3  \right].
\end{equation}
The expression above is exact.  However, to simplify the calculation, we approximate  $ 	(a\sqrt{\epsilon_1})''/ a \sqrt{\epsilon_1} \simeq 2/\eta^2$, i.e. we have used expression \eqref{appz} and only kept  zero-order terms in the HFF.  The general solution of \eqref{eqMSfourier} is then
\begin{eqnarray}\label{solMS}
	\mM_k (\eta) &=& C_1  \frac{3^{1/4}}{\sqrt{2k}} \left(1 - \frac{\sqrt{3}i}{k \eta}\right) e^{-i k \eta/\sqrt{3}}\nn
&+& C_2  \frac{3^{1/4}}{\sqrt{2k}} \left(1 + \frac{\sqrt{3}i}{k \eta}\right) e^{+i k \eta/\sqrt{3}}.
\end{eqnarray}
For the moment, let us fix the constants $C_1=1$, $C_2 =0$ and save the discussion about the  initial conditions for the end.  Therefore, the amplitude of the  comoving curvature perturbation  is
\begin{equation}\label{Rkamplitud}
	| \mR_k (\eta) |^2 = \frac{3^{1/2}}{4k M_P^2 a^2 \epsilon_1 } \left(   1 + \frac{3 \mH^2}{k^2}    \right) ,
\end{equation}
where we have used once again the approximation $\mH \sim 1/\eta$. For super-Hubble modes $k/\sqrt{3} \ll \mH$, i.e. modes whose proper wavelength becomes larger than the Hubble radius during inflation, we obtain
\begin{equation}\label{RKSH}
	k^3	| \mR_k (\eta) |^2_{SH} \simeq \frac{3^{3/2} H^2}{4 M_P^2  \epsilon_1 } ,
\end{equation}
where we have used that $\mH = a H$. The dimensionless primordial power spectrum for super-Hubble modes  is given as $\mP_\mR(k)  \equiv 	k^3	| \mR_k (\eta) |^2_{SH} /(2 \pi^2)$. Henceforth, the power spectrum corresponding to the comoving curvature perturbation $\mR_k$ is,
\begin{equation}\label{PS}
	\mP_\mR(k) =  \frac{3^{3/2} H^2}{8 \pi^2 M_P^2  \epsilon_1 }.
\end{equation} 
 The obtained spectrum \eqref{PS} is manifestly scale invariant, but this is because we have used Eq. \eqref{appz} and retained only the zero-order term in the HFF. Taking into account higher order terms, will yield the usual scale dependence of $\mP_\mR(k) $ as characterized by the spectral index $n_s = 1- 2 \epsilon_1 - \epsilon_2$. 
 
Thus, we have shown that the same primordial power spectrum as in the standard inflationary scenario, can be obtained in our model as well.   Furthermore,  this result is independent of the particular function $Q$. The reason being is that, as we have shown in this section,  the theory of cosmological perturbations in UG,  along with assumptions (i) and (ii), yields exactly the same set of  equations  for the perturbations as in standard inflation [see Eqs. \eqref{dEFE}]. In particular, one retrieves the well known Mukhanov-Sasaki equation for the comoving curvature perturbation. Therefore, the standard prediction for the primordial spectrum is obtained.   The diffusion term $Q$ only affects the dynamics of the background.  In other words,  if $Q$ dominates over the energy density of the ultra-relativistic fluid, and satisfies Eq. \eqref{condicionQinf2}, then it can characterize an inflationary phase equivalent to slow-roll inflation.  In this way, the predicted spectrum will possess the correct amplitude and shape.


In this final part of the section, we return to the subject of initial conditions in Eq. \eqref{solMS}.  Up to this point our analysis has been classical.  In view  that the MS equation is exactly the same as in standard inflation, and that $\mR$ is a gauge-invariant quantity also in UG, we can adopt $\mM = \mR z$ as the canonical quantum field variable.  The only difference, with traditional inflation models, is that here $\mM$ is a combination of the metric perturbations $\psi$, $B$ and the velocity potential $v$ (see Eq. \eqref{Rcomov}),  while in slow-roll inflation models, the definition of  $\mM$ involves the combination  $\psi$ and $\delta \varphi$ (where $\varphi$ represents the inflaton).  One can write then an action up to second order in the $\mM$ variable that yields Eq. \eqref{MSmov}, i.e.
\begin{equation}\label{action0}
	\delta^2 S = \frac{1}{2} \int d^4 x \left( \mM'^2 - \frac{1}{3} (\nabla \mM)^2   + \frac{z''}{z} \mM^2  \right).
\end{equation}
This action is completely analogous to the action found for perturbations of standard hydrodynamical matter in the cosmological context using GR [see Eq. (10.62) of \cite{mukhanov92}].  We can then apply standard methods of quantization for the Mukhanov-Sasaki variable,   in which the Fourier modes of the quantum field, i.e.
\begin{equation}\label{quantumM}
	\hat \mM (\eta,\x) = \int \frac{d^3 k}{(2 \pi^{3/2})} [\hat a_\nk \mM_k (\eta) e^{-i  \nk \cdot \x} + \text{c.c.}]
\end{equation}
satisfy   Eq. \eqref{eqMSfourier}.  The annihilation operators $\hat a_\nk$ and $\hat a_\nk^\dagger$ obey the standard commutation relations. At this point, one must make a choice regarding the vacuum state. As usual, we can select the Bunch-Davies vacuum, leading to the initial conditions selected previously in Eq. \eqref{solMS}, i.e. $C_1 = 1, C_2 =0$.  The power spectrum of the scalar perturbations, is then given by the relation
\begin{eqnarray}\label{quantumPS}
	\int \frac{dk}{k} \mP_\mR (k) &=& \frac{1}{z^2} \int \frac{d^3(\x-\y)}{(2\pi^3)} \bra 0 | \hat \mM (\eta,\x) \hat \mM (\eta,\y) | 0 \ket \nn
	&\times& \exp[-i \nk \cdot (\x-\y) ],
\end{eqnarray}
where $| 0 \ket$  is the vacuum state defined as $\hat a_\nk |0 \ket=0 $ for all $\nk$. The scalar power spectrum obtained is then $\mP_\mR(k)  = 	k^3	| \mR_k (\eta) |^2 /(2 \pi^2)$. 

The selection of a combination of metric perturbations and the velocity potential $v$ as a quantum variable, has been considered previously. For instance, in Refs. \cite{Wald2002,Bengochea2014} this same quantization scheme was used. However, in those works, no inflationary background was present, instead the  universe expansion was driven by standard hydrodynamical matter in GR resulting in a non-accelerated expansion.  As such, in those works a $k$ dependence in $\mP_\mR (k)$ was obtained, different from a nearly scale invariant spectrum unless some additional assumptions were made.  Note also that the quantization procedure presented here is very similar to the conventional one, but differs significantly from that of Ref. \cite{aperez2021}, where fluctuations of the  Higgs field were subjected to quantization within the semiclassical gravity approximation.

The main result of this section is that the known results of cosmological perturbation in GR involving the comoving perturbation $\mR$ variable are recovered completely under the assumptions (i) and (ii) introduced before. In addition,  the method involving the quantization of the MS field variable is exactly equivalent as well.  However, in our model, the inflationary phase is driven exclusively by the diffusion term $Q$, and the matter content consists of a conventional fluid modeled as  pure radiation, i.e. no extra fundamental field, like the inflaton, was postulated to exist.

\section{Post-inflationary dynamics and the present value of the cosmological constant}\label{sec_CC}

In the preceding two sections we have shown that a diffusion (homogeneous) term $Q$ can account for an early epoch of accelerated expansion equivalent to slow-roll inflation.  In particular, if the matter content during such a phase behaves as pure radiation, then the primordial power spectrum, associated to the density perturbations of the fluid, has exactly the same amplitude and shape as in traditional models of inflation. 

In this section, we are interested in analyzing the dynamics of the diffusion term $Q$ after inflation has ended. For the sake of doing a quantitative analysis, we assume  ansatz \eqref{Qsol} throughout this section.  However, we will leave the full statistical analysis using recent observational data for  future research. Instead, here we will be more concerned with the orders of magnitude of some cosmological observables.  We acknowledge that the precision of cosmological data currently available is high enough to disprove any significant departure from the standard $\Lambda$CDM cosmological model.  Therefore, the results obtained  in this section are very preliminary, and we do not take them as indicating that our proposal is a complete and finished alternative to the $\Lambda$CDM model.  Instead, our position is to investigate whether the values obtained  for some important cosmological observables are realistic or not.

After the inflationary period has ended, it is natural to consider the universe matter content still as  single ultra-relativistic fluid (i.e. as pure radiation).  However, below  the electroweak symmetry breaking temperature, $T \leq 100$ GeV, the gauge bosons of the weak interactions receive masses through the Higgs mechanism. Thus, below this temperature,  we approximate the matter content of the universe as consisting of two type  of components: presureless dust $w=0$ also called matter for simplicity, and radiation $w=1/3$. 

The continuity equation \eqref{rhomov}, using $N$ as the dynamical variable,  is then
\begin{equation}\label{key}
	\sum_i \rho_i,_N +  3 \rho_i (1 + w_i) + Q,_N =0,
\end{equation}
where  $i = m,\gamma$ denotes  matter and radiation respectively.  

The next important temperature scale  is  $T_\eq \simeq 0.5$ eV where the matter-radiation equality occurs, that is when there is equal abundance of matter and radiation in the universe;  we denote the energy density at this stage as $\rho_\eq$.   During the radiation dominated epoch, $ T> T_\eq$,  we consider that $\rho_\gamma$ remains  coupled directly to the diffusion term $Q$ (as same  as during inflation), while $\rho_m$ behaves free of such influence. In particular, for $T > T_\eq$, the energy densities $\rho_m$ and $\rho_\gamma$ satisfy
\begin{equation}\label{rhomovrad}
\rho_\gamma,_N +  4 \rho_\gamma  + Q,_N =0,  \qquad \rho_m,_N + 3 \rho_m = 0.
\end{equation}
Given that  Eq. \eqref{Qsol} is the ansatz for $Q$, the solution $\rho_\gamma$ is exactly the same as Eq. \eqref{rhosol}, while the solution for the matter density is
\begin{equation}\label{rhomrad}
	\rho_m (N) = \frac{\rho_\eq}{2} e^{-3 (N-N_\eq)} \qquad \text{for} \quad  T > T_\eq.
\end{equation}

In the opposite regime, during the matter dominated epoch $T<T_\eq$, we consider that the diffusion term $Q$ is now coupled to $\rho_m$, and $\rho_\gamma$ becomes free of this term.   In other words, we are assuming that, since the very beginning of the universe's expansion, $Q$ always couples to the dominant type of matter. In particular, for $T < T_\eq$, the energy densities $\rho_m$ and $\rho_\gamma$ satisfy
\begin{equation}\label{rhomovmat}
	\rho_m,_N +  3 \rho_m  + Q,_N =0,  \qquad \rho_\gamma,_N + 4 \rho_\gamma = 0.
\end{equation}
The solution $\rho_\gamma$ in this regime is
\begin{equation}\label{rhogammasolmat}
	\rho_\gamma (N) = \frac{\rho_\eq}{2} e^{-4 (N-N_\eq)} \qquad \text{for} \quad  T < T_\eq.
\end{equation}
and $\rho_m(N)$ is
\begin{eqnarray}\label{rhomsolmat}
		\rho_m (N) &=&  \frac{\rho_\eq}{2} e^{-3 (N-N_\eq)} \nn
	&  & - e^{-3 N} \int_{N_\eq}^N e^{3 \bar{N}} Q,_{\bar N} d\bar N  \qquad \text{for} \quad  T < T_\eq.  \nn
\end{eqnarray}
Given the ansatz in Eq. \eqref{Qsol} corresponding to $Q$, one can find $\rho_m(N)$ from \eqref{rhomsolmat}. 

The dynamics of $\rho_\gamma$, $\rho_m$ and $Q$, enable us to find an estimate of their corresponding order of magnitude at the present epoch. Furthermore, we recall  that the diffusion term $Q$, as given  in Eq. \eqref{Qsol},  involves two free parameters: $\alpha$ and the total number of e-folds that inflation lasts $N_f$.  Thus, once these two parameters are fixed,  secondary parameters are settled in terms of these two.    As explained in Sec. \ref{sec_inf},  we  choose  $N_f = 300$ and $\alpha \simeq 0.03$.

We now focus on the radiation energy density $\rho_\gamma$.  In this case, we match solutions \eqref{rhosol} and \eqref{rhogammasolmat} at $T_\eq \simeq 0.5$ eV. For different values of $\alpha$, the precise value $N_\eq$ changes accordingly.  We have found that for $\alpha = 0.027$, which corresponds to $N_\eq =357.3 $,  and for $N_0 = 362.3$  (where $N_0$ corresponds to the number of e-folds at present epoch $a_0 = e^{N_0} a_\ini$), the predicted radiation energy density today is $\rho_\gamma (N_0) \simeq 10^{-124} M_P^4$, or equivalently,  a  radiation energy scale of $10^{-4}$ eV.  This order of magnitude is consistent with the present background radiation temperature as measured by recent experiments, e.g. \cite{Planck18a,Planck18c}. 

In the case of the matter energy density,  $\rho_m (N)$ in \eqref{rhomsolmat}, we have found that, for the same value of the parameters as before: $\alpha = 0.027$, $N_f =300$, $N_\eq =357.3 $ and $N_0 = 362.3$, the predicted energy scale today of the matter density is $\rho_m^{1/4} (N_0) \simeq 10^{-30} M_P$ (we have resolved the integration in \eqref{rhomsolmat} numerically).  This energy scale is very close to the order of magnitude observed today, which can be checked as follows. From the definition  $ \Omega_{m,0} \equiv \rho_{m,0} / \rho_{c,0}$, where $\rho_{c,0} \equiv 3 H_0^2 M_P^2$ is the present value of the critical energy density.  And, using the following values (obtained from \cite{Planck18a,Planck18c}): $\Omega_{m,0} = 0.315$, $H_0 = 5.9776 \times 10^{-61} M_P$ (which corresponds to $H_0 = 67.3$ km/s/Mpc), one finds  the energy scale $\rho_{m,0}^{1/4} = 7.62 \times 10^{-31} M_P$, which is pretty similar to the order of magnitude obtained in our model.

We now proceed to analyze  the diffusion term $Q$ as given in Eq. \eqref{Qsol}. We remind the reader that  $Q(N)$ was constructed in Sec. \ref{sec_inf}, and is able to generate an inflationary phase with the desired characteristics to be considered as realistic as possible.  In particular,  the inflationary regime lasts long enough, and exhibits a graceful exit towards the radiation dominated era.  The specific function $Q(N)$ in \eqref{Qsol} represents just a possible ansatz (compatible with natural initial conditions), but once we have chosen that ansatz, it remains the same throughout the universe's history.  

\begin{figure}
	\centering
	\includegraphics[scale=1.0]{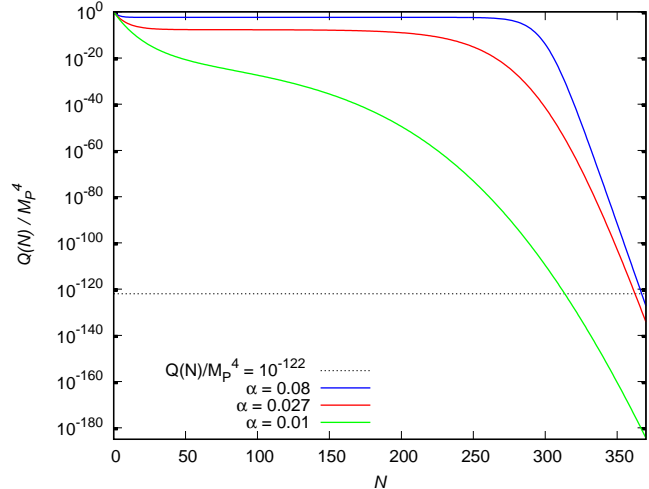}
	\caption{ The diffusion term $Q (N)$ in Eq. \eqref{Qsol} for different values of $\alpha$ and assuming a total duration of inflation of $N_f = 300$.  The  chosen values of $\alpha$ serve to provide a visualization of the effects in varying such a parameter.  In fact, for $\alpha = 0.027$ and $N_0 = 362.3$, we observe that $Q(N_0) \simeq 10^{-122} M_P^4$, which is pretty close to the order of magnitude of the cosmological constant as measured today \cite{Planck18a,Planck18c}. }
	\label{fig_QN}
\end{figure}

In Fig. \ref{fig_QN}, we have plotted the complete evolution of  the diffusion term $Q(N)$ for some values of $\alpha \simeq 0.03$. We note that for $\alpha = 0.027$ and $\alpha = 0.08$, $Q$ remains approximately constant between $N=0$ and $N_f = 300$, which corresponds to the inflationary phase.  That behavior is consistent with Eq. \eqref{Qinftipico}, in which $Q$ was estimated to be a constant with a corresponding value similar to the characteristic energy scale of inflation. On the other hand, for $\alpha=0.01$, the diffusion term fails to remain constant during inflation. In fact, $Q$ decreases  for several orders of magnitude between $N=0$ and $N_f = 300$.  Thus, for $\alpha = 0.01$, the corresponding diffusion term \eqref{Qsol}, does not characterize a realistic inflationary phase. That can also be concluded from Fig.  \ref{fig_R}-right (as was discussed in Sec. \ref{sec_inf}).  There, we observe that for $\alpha = 0.01$, the function $Q_{,N}/Q$ does not satisfy Eq. \eqref{condicionQinf2}, i.e. $Q_{,N}/Q$  never approaches zero enough from the bottom half plane.

After the end of inflation, we observe in Fig. \ref{fig_QN} that $Q$ decays steadily.  In particular, we have found that for the same values of the parameters as before: $\alpha = 0.027$, $N_f =300$ and $N_0= 362.3$, one obtains 
\begin{equation}\label{Qhoy}
	Q(N_0) \simeq 10^{-122} M_P^4,
\end{equation}
 which is remarkably very similar to the order of magnitude corresponding to the observed value of the cosmological constant today.  Specifically, the cosmological constant $\Lambda$ is expressed as  $\Lambda = 3 H_0^2 \Omega_{\Lambda,0}$, using the values $\Omega_{\Lambda,0}  = 0.6889$ and   $H_0 = 5.9776 \times 10^{-61} M_P$ (obtained from \cite{Planck18a,Planck18c}), one finds the order of magnitude $\Lambda \simeq 10^{-122} M_P^2$  (recall that in the units used in the present work $\Lambda = Q/M_P^2$).

Finally, it is also worth mentioning that the idea of a decaying cosmological constant  has been considered before in the literature \cite{sudarskyPRL1,sudarskyPRL2,CCvar0,CCvar1,CCvar2,CCvar3,CCvar4}.  An interesting scenario is the Early Dark Energy (EDE) model \cite{EDEPRL,EDE0,EDE1} which has been proposed as a possible solution to the so called Hubble tension.  A particular realization of this model \cite{EDE4,EDEPRL}, consists of  a scalar field $\varphi$ with  a potential $V (\varphi) \propto (1 - \cos[\varphi/f ])^n$ . At early times, the field is ``frozen'' and acts as a cosmological constant. Nonetheless,  when $H$ decreases below some value, at a critical redshift $z_c+1 =a_0/a_c$ (where $a_0$ denotes the value of the scale factor today), the fluid begins to oscillate, effectively behaving as a fluid with an equation of state $w_n = (n-1)(n+1)$. The precise details of the model can be found elsewhere (see e.g. \cite{EDEPRL,EDE2,EDE3}). The aspect that concern us here is that the evolution of $\Lambda$ in such a model can be parameterized as follows:
\begin{equation}\label{lambdaEDE}
	\Lambda_\text{EDE} (N) = \frac{2 \Lambda(N_c)}{1+ \exp[ \beta_n (N-N_c)  ]},
\end{equation}
where $\beta_n \equiv 3(w_n+1)$,   $N_c$ corresponds to the number of e-folds at the critical redshift $z_c$, that is, $a_c = e^{N_c} a_\ini$. 
 In Ref.  \cite{EDEPRL}, it was found that for $w_n = 1/3$, the critical redshift best fit value to their data set is $\log_{10} (a_c/a_0)  = -3.728$, which corresponds to $N_c = N_0 - 8.584$. This found value also seems to alleviate the Hubble tension \cite{Dainotti2021,Dainotti2022} without introducing significant changes to the rest of the standard cosmological parameters.

\begin{figure}
	\centering
	\includegraphics[scale=1.0]{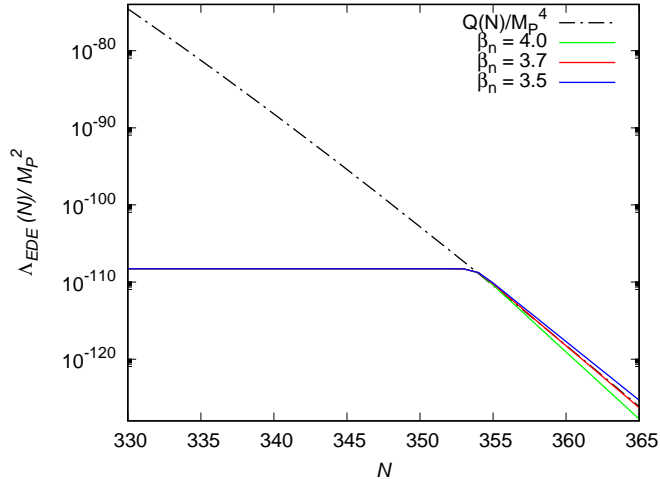}
	\caption{ The function $\Lambda_\text{EDE} (N)/M_P^2$ in \eqref{lambdaEDE} and three distinct values of $\beta_n$. We have included also the diffusion term $Q (N)/M_P^4$  (dashed line) using: $\alpha=0.027$, $N_f = 300$ and $N_0=362.3$.  We have matched the two functions at $N_c = 353.7 $.  For $\beta_n = 3.7$ both functions decay pretty similar after $N_c$. }
	\label{fig_QEDE}
\end{figure}

In order to compare $Q(N)$ and $\Lambda_\text{EDE} (N) $, we match the two functions at $N_c$; this is, $\Lambda(N_c) = Q(N_c)/M_P^2$, where $N_c = N_0 - 8.584$. For the function $Q(N)$ we repeat the values: $\alpha = 0.027$, $N_f = 300$ and $N_0=362.3$.  With these assumptions, in Fig. \ref{fig_QEDE}  we plot $Q(N)/M_P^4$ and $\Lambda_\text{EDE} (N) /M_P^2$ for three different values of $\beta_n$.  The case $\beta_n =4$ corresponds to $w_n = 1/3$, i.e. $\Lambda_\text{EDE} (N) $ decays in the same manner as radiation.  On the other hand,  we observe that for $\beta_n = 3.7$, the dynamical behavior of the two functions, $Q(N)/M_P^4$ and $\Lambda_\text{EDE} (N)/M_P^2 $, is very similar.  

The former result is very encouraging since it opens the possibility to include all the advantages of the EDE scenario in our model; particularly, the resolution of the Hubble tension. Furthermore, in our model, we would not need  to  postulate any external scalar field, which usually implies the introduction of additional parameters characterizing its potential. 

The results obtained in this section show that, for the particular set of parameters: $\alpha = 0.027$, $N_f = 300$, 
the order of magnitude corresponding to  $\rho_\gamma (N_0), \rho_m (N_0), Q(N_0)$ and $H(N_0)$  is compatible with the order of magnitude realized from observational data. However, as we have argued at the beginning of this section, the analysis presented here is just  preliminary and much work is still required.

\section{Conclusions}\label{conclusions}

In this work, we have explored the consequences of assuming an inflationary phase generated by a homogeneous \textit{diffusion term} $Q$. This term, is identified with the source of the violation of the energy-momentum conservation $\nabla^\mu (T_{\mu \nu} -Q g_{\mu \nu})=0$,  a feature that is not prohibited within the unimodular gravity theory due to its invariance under volume preserving diffeomorphisms.  The dominant type of matter assumed during this inflationary phase is standard hydrodynamical matter consisting of a perfect fluid with an equation of state of pure radiation ($p=\rho/3$).  In this way, we did not have to postulate the existence of the inflaton, which automatically implies to assume a particular shape of its potential and  initial conditions.  

We have found the conditions for a generic  $Q$, required to reproduce a realistic inflationary phase. Moreover, for a parameterization of inflation, expressed in terms of the Hubble flow functions, we have shown a method to construct a corresponding diffusion term $Q$, that can recreate the main features of traditional inflation (e.g. slow-roll inflation). In particular, the inflationary phase lasts long enough and ends in a graceful manner. Therefore, the results obtained are quite general.  Moreover, the predicted primordial spectrum of density perturbations, corresponding to the radiation fluid, has the same amplitude and shape as the traditional one.  This is because the modes obey the well known Mukhanov-Sasaki equation and can be subjected to the usual quantization procedure.  The former result is independent of the particular form of $Q$, as long as it satisfies the conditions to characterize inflation.

In our model, as the universe evolves through the radiation and matter epochs, the diffusion term $Q$ decays steadily until it reaches the present epoch. For the ansatz \eqref{Qsol} constructed in Sec. \ref{sec_inf}, and for some particular values of the free parameters, the order of magnitude of $Q$ estimated today is consistent with the observed value of the cosmological constant.  In addition, the estimated order of magnitude corresponding to $H_0$,  radiation and matter energy densities in our model matches  the observable range. 

On the other hand,  in order to present a complete and finished alternative to the standard $\Lambda$CDM cosmological model, much work is still required. In particular, we have not discussed the micro-physics that might be responsible for the diffusion term $Q$. So there is still the possibility that trans-Planckian or quantum gravity effects might alter $Q$ in such a way that inflation cannot be sustained.  Additionally,  even at the phenomenological level, a full statistical analysis using  high precision cosmological data is needed.  However, the preliminary results obtained in this work open an interesting possibility to acquire a unified picture of the primordial and present accelerated expansion of the universe.

\begin{acknowledgments}
The author thanks the anonymous referees for their comments and suggestions. G.L. is supported by CONICET (Argentina);  he also acknowledges support from the following project grants: Universidad Nacional de La Plata I+D  G175 and  PIP 11220200100729CO  CONICET (Argentina).

\end{acknowledgments}



\bibliography{bibliografia}
\bibliographystyle{apsrev}

\end{document}